\documentclass[a4paper,10pt]{article}

\usepackage{amsmath}
\usepackage{amsthm}
\usepackage{amssymb}
\usepackage{eepic}
\usepackage{amscd}

\newtheorem{theorem}{Theorem}[]
\newtheorem{lemma}[theorem]{Lemma}

\newtheorem{proposition}[theorem]{Proposition}

\newtheorem{example}[theorem]{Example}

\theoremstyle{definition}
\newtheorem{definition}[theorem]{Definition}

\newtheorem{remark}[theorem]{Remark}

\newlength{\fuyasu}
\theoremstyle{definition}

\newcommand{\gehn}{\ensuremath{\mathfrak{g}_n}}

\newcommand{\ot}{\otimes}

\newcommand{\Z}{{\mathbb Z}}

\newcommand{\Pb}{{\bar{P}}}

\newlength{\BallWidth} 
\newcommand{\batu}[1]{
{\setlength{\unitlength}{10pt}
	\put(-1,0.3){\vector(1,0){2.4}}
	{\thicklines\put(0,1.5){\vector(0,-1){2.5}}}
}}

\newcommand{\yajirusiK}[1]{
$K_{#1}$\;
{\setlength{\unitlength}{10pt}
	{
	\qbezier(0.5,1.5)(0,1)(0,0)
	\qbezier(0,0)(0,-0.5)(0.5,-1)
	\put(0.5,-1){\vector(3,-4){0}}
	}
}}

\newcommand{\batten}[5]{%
\begin{picture}(40,40)(-20,-20)
	\put(-10,0){\vector(1,0){20}}
	\thicklines
	\put(0,10){\vector(0,-1){20}}
	\put(-11,0){\makebox(0,0)[r]{$#1$}}
	\put(0,11){\makebox(0,0)[b]{$#2$}}
	\put(0,-11){\makebox(0,0)[t]{$#3$}}
	\put(11,0){\makebox(0,0)[l]{$#4$}}
\end{picture}
}

\title{Factorization, reduction and embedding \\
in integrable cellular automata }

\author{
Atsuo Kuniba\thanks{
Institute of Physics, University of Tokyo, Komaba, Tokyo 153-8902, Japan},
Taichiro Takagi\thanks{
Department of Applied Physics, National Defense Academy,
Kanagawa 239-8686, Japan}, and
Akira Takenouchi$^*$}

\date{}
\begin{document}
\maketitle

\begin{abstract}
Factorized dynamics in 
soliton cellular automata with quantum group symmetry
is identified with a motion of particles and anti-particles
exhibiting pair creation and annihilation.
An embedding scheme is presented showing that 
the $D^{(1)}_n$-automaton contains, as certain subsectors, 
the box-ball systems and all the other automata associated 
with the crystal bases of non-exceptional affine Lie algebras.
The results extend the earlier ones to 
higher representations by a certain reduction and 
to a wider class of boundary conditions.
\end{abstract}

\section{Introduction}\label{sec:1}

In \cite{HKT1}, a class of one dimensional 
cellular automata with quantum group symmetry were introduced.
They are associated with the crystal bases \cite{KKM} 
of non-exceptional affine Lie algebras 
$A^{(2)}_{2n-1}, A^{(2)}_{2n}, B^{(1)}_n, C^{(1)}_n, D^{(1)}_n$ 
and $D^{(2)}_{n+1}$.
Since then, 
(i) configurations that behave as solitons are 
extracted \cite{HKT1}, 
(ii) scattering of solitons is determined 
in terms of combinatorial $R$ by utilizing the quantum group symmetry
\cite{HKOTY}, 
(iii) time evolution under a sufficiently large carrier 
is factorized into a product of Weyl group operators \cite{HKT2}, 
(iv) the resulting factorized dynamics is described by an 
elementary algorithm using particles and anti-particles \cite{HKT3},
(v) tropical $R$ for geometric crystals is formulated \cite{KOTY1}, 
which yileds the evolution equation of the automata under 
ultradiscretization \cite{TTMS},
(vi) the tropical $R$ is bilinearized in terms of tau functions 
of the DKP hierarchy \cite{KOTY2}, allocating the
automata in classical integrable systems.

Through the developments these automata have now been 
understood at a level comparable with the 
well known box-ball systems \cite{TS,T,TM}
corresponding to $A^{(1)}_n$ \cite{HHIKTT,FOY}.
The factorization of the time evolution in this case was known 
as the decomposition into a finer process where one only moves 
the balls with a fixed color.

The states of our automata have the form (\ref{eq:W}), 
where $a$ specifies the boundary condition 
and $l = (l_i)_{i \in \Z}$ denotes the `capacities of the 
aligned boxes' in the terminology of the box-ball systems and 
`magnitude of local spins' in the terminology of solvable lattice models.
Throughout the paper the automata with $(\forall l_i\!=\!1)_{i \in \Z}$
and general $(l_i)_{i \in \Z}$ will be called basic and 
inhomogeneous, respectively.

The goal of this paper is to provide the full generalization of 
the above result (iv) which was limited to 
the basic automata with the boundary condition $a=1$.
Our approach involves three essential 
ingredients; {\em factorization} (section \ref{sec:2}), 
{\em reduction} (section \ref{sec:3}) and 
{\em embedding} (section \ref{sec:4}).
Factorization of the time evolution into the Weyl group operators 
had been proved for any inhomogeneous automata associated with
$A^{(1)}_n, A^{(2)}_{2n-1}, A^{(2)}_{2n}, 
B^{(1)}_n, C^{(1)}_n, D^{(1)}_n$ and $D^{(2)}_{n+1}$ in \cite{HKT2}.

In sections \ref{sec:2} and \ref{sec:3} we concentrate on 
the $D^{(1)}_n$ case and the former 
is devoted to the quotation of the result in \cite{HKT2}.
Section \ref{sec:3} contains our main Theorem \ref{th:main},
which claims that under any boundary condition $a$, 
the factorized dynamics of the inhomogeneous $D^{(1)}_n$-automaton 
admits a reduction to the basic one except inserting 
a certain `reshuffling operator' $Q$.
See Examples \ref{ex:Td4} and \ref{ex:kactionD}.
It integrates the crystal basis theory 
on the time evolution 
into the simple diagrams (\ref{eq:batt}), which represent a motion
of particles and anti-particles
that undergo pair creation and annihilation.
Until this point we achieve double simplifications of the inhomogeneous 
$D^{(1)}_n$-automaton;
first by the factorization with respect to `color' of the particles
and second by the reduction to the basic case up to reshuffling.
They make the time evolution a handy enough calculation and suit programming 
due to the locality of the diagrams (\ref{eq:batt}).

In section \ref{sec:4} we explain the embedding, by which 
the $D^{(1)}_n$-automaton can be regarded as a `master automaton' 
that contains, as certain subsectors, 
all the other inhomogeneous automata associated with 
$A^{(1)}_n, A^{(2)}_{2n-1}, A^{(2)}_{2n}, 
B^{(1)}_n, C^{(1)}_n$ and $D^{(2)}_{n+1}$.
In a sense this is the third simplification 
that unites the number of results obtained previously.
Our embedding scheme is a synthesis of the 
preceeding ones in \cite{HKOTY} and \cite{KOTY2}.

In the box-ball system associated with $A^{(1)}_{n-1}$, 
all the colors are equivalent under $\Z_n$ symmetry.
Such a symmetry is absent in the relevant crystal graph for 
$D^{(1)}_n$ in Figure \ref{fig:B} in appendix \ref{app:a}.
This motivates us to study the automaton under 
a general boundary condition specified by any point 
$a$ in the crystal graph.
Since time evolutions under any boundary conditions are 
connected by a similarity transformation 
(see Definition \ref{def:calT}),
qualitative features of the automata remains the same.
However, solitons show up in various guises 
depending on $a$.
Section \ref{sec:soliton} contains the list of such configurations 
under the general boundary condition.

We defer the technical details 
to section \ref{sec:5} and appendix \ref{app:b},
where a proof of Theorem \ref{th:main} is outlined.
The key is to identify the crystal theoretic 
Weyl group operator $S_i$ \cite{K} with the  
particle motion operator $K_j$ via a certian gauge transformation 
as in Proposition \ref{pr:main} and Lemma \ref{lem:sk}.
We note that for the box-ball systems, a reduction to the basic case
was also known in \cite{HHIKTT,F}.

\section{\mathversion{bold}
Factorization in inhomogeneous $D^{(1)}_n$-automaton}\label{sec:2}
For a positive integer $k$,
let $B_k$ be the crystal for $k$-fold symmetric tensor
representation of $D^{(1)}_n$ ($n \ge 4$) \cite{KKM}.
Each element of $B_k$ is written as
$\vec{x}=(x_1,\ldots,x_n,x_{-n},\ldots,x_{-1})$, where
$x_i$'s are non-negative integers that add up to $k$ and satisfy the 
condition $x_n x_{-n}=0$.
Let $J=\{ \pm 1,\ldots,\pm n \}$.
For $a \in J$ we let $\vec{a} \in B_k$ denote the unique
element characterized by $x_a = k$.
(We suppress the dependence on $k$.)
Let $l=(l_i)_{i\in \Z}$ be an array of positive integers.
Given $a \in J$ and $l = (l_i)_{i\in \Z}$, 
recall the inhomogeneous $D^{(1)}_n$-automaton \cite{HKT2} having the 
capacity $l$ and the prescribed boundary condition $a$.
It is a discrete dynamical system over the set
\begin{equation}\label{eq:W}
W[a]_l = \{ (\ldots, \vec{b}_i, \vec{b}_{i+1},\ldots) \in
\cdots \times B_{l_i} \times B_{l_{i+1}} \times \cdots \mid
\vec{b}_i = \vec{a}
\text{ for } \vert i \vert \gg 1 \},
\end{equation}
where the product $\times$ of sets 
actually means the tensor product of the 
crystals $B_{l_i}$.

Let
$S_0, \ldots, S_n$ be the Weyl group operators \cite{K} acting on  
$\coprod_{a \in J} W[a]_l$.
They satisfy the Coxeter relations, which 
include in particular $S^2_i = id$, 
$S_0S_1 = S_1S_0$ and $S_nS_{n-1} = S_{n-1}S_n$.
We introduce the operator $\sigma^\Delta$ that acts on 
$\boldsymbol{b} 
= (\ldots, \vec{b}_i, \vec{b}_{i+1},\ldots) \in \coprod_{a \in J}W[a]_l$
as
$\sigma^\Delta(\boldsymbol{b})
=(\ldots, \sigma(\vec{b}_i), \sigma(\vec{b}_{i+1}),\ldots)$, 
where $\sigma(\vec{b}_i)$ is obtained {}from 
$\vec{b}_i = (x_1,\ldots, x_n,x_{-n},\ldots, x_{-1}) \in B_{l_i}$ 
by the simultaneous interchanges $x_1 \leftrightarrow x_{-1}$ and 
$x_n \leftrightarrow x_{-n}$.
It induces the involutive automorphism
$\sigma^\Delta S_0 \sigma^\Delta = S_1$, 
$\sigma^\Delta S_{n-1} \sigma^\Delta = S_n$ and 
$\sigma^\Delta S_i \sigma^\Delta = S_i$ for $i\neq 0, n$.

In terms of these operators 
the time evolution $\mathcal{T}_a: W[a]_l \rightarrow W[a]_l$ 
of our $D^{(1)}_n$-automaton is defined as follows.
\begin{definition}\label{def:calT}
\begin{align*}
&\mathcal{T}_1 = \sigma^\Delta 
S_0S_2S_3\cdots S_{n-2}S_nS_{n-1}\cdots S_2S_0,\\
&\mathcal{T}_a = S_{a-1}^{-1} \mathcal{T}_{a-1} S_{a-1}
\quad (\mbox{for} \quad 2 \leq a \leq n)\\
&= \begin{cases}
\sigma^\Delta 
S_aS_{a+1} \cdots S_{n-2}S_nS_{n-1}\cdots S_2S_0S_1\cdots S_{a-1} &
2 \le a \le n-1,\\
\sigma^\Delta 
S_{n-1}S_{n-2} \cdots S_2S_0S_1 \cdots S_{n-1} & a=n,
\end{cases}\\
&\mathcal{T}_{-n} = S_n^{-1} \mathcal{T}_{n-1} S_n 
= \sigma^\Delta 
S_nS_{n-2} \cdots S_2S_0S_1 \cdots S_{n-2}S_n,\\
&\mathcal{T}_{-a} = S_a^{-1} \mathcal{T}_{-a-1} S_a
\quad (\mbox{for} \quad 1 \leq a \leq n-1)\\
&= \begin{cases}
\sigma^\Delta 
S_{a-1}S_{a-2} \cdots S_2S_0S_1 \cdots S_{n-2}S_n
S_{n-1}\cdots S_a & 2 \le a \le n-1,\\
\sigma^\Delta 
S_1S_2S_3\cdots S_{n-2}S_nS_{n-1}\cdots S_2S_1 & a = 1.
\end{cases}
\end{align*}
\end{definition}
These operators can be extended to $\coprod_{a \in J}W[a]_l$, 
where $\mathcal{T}_a = (\mathcal{T}_{-a})^{-1}$ is valid for any 
$1 \le a \le n$. 
See also (\ref{eq:calTexp}) and (\ref{eq:iprop}) for 
a property of the Weyl group elements.
\begin{example}\label{ex:calT}
For $n=4$, $\mathcal{T}_a$'s read
\begin{align*}
&\mathcal{T}_1 = \sigma^\Delta S_0S_2S_4S_3S_2S_0,\\
&\mathcal{T}_2 = \sigma^\Delta S_2S_4S_3S_2S_0S_1,\\
&\mathcal{T}_3 = \sigma^\Delta S_4S_3S_2S_0S_1S_2,\\
&\mathcal{T}_4 = \sigma^\Delta S_3S_2S_0S_1S_2S_3,\\
&\mathcal{T}_{-4} = \sigma^\Delta S_4S_2S_0S_1S_2S_4,\\
&\mathcal{T}_{-3} = \sigma^\Delta S_2S_0S_1S_2S_4S_3,\\
&\mathcal{T}_{-2} = \sigma^\Delta S_0S_1S_2S_4S_3S_2,\\
&\mathcal{T}_{-1} = \sigma^\Delta S_1S_2S_4S_3S_2S_1.
\end{align*}
\end{example}
The automaton was originally introduced in \cite{HKT1} 
for $\forall l_i = 1$ and $a=1$ case, where the time evolution 
$\mathcal{T}_1$ was defined by using the combinatorial $R$.
It was extended to general inhomogeneity $l$ and boundary condition $a$
in \cite{HKT2}, where it was proved that the $\mathcal{T}_a$ 
is factorized as above into the Weyl group operators.
Due to the boundary condition, it can be shown that 
all the $S_i$ act actually as $S_i = e_i^\infty$.
The expressions in Definition \ref{def:calT} do not depend on 
$l$ explicitly.
The basic $D^{(1)}_n$-automaton corresponding to the situation
$\forall l_i = 1$ will play a fundamental role 
in the sequel.

\section{\mathversion{bold}
Reduction to basic $D^{(1)}_n$-automaton}\label{sec:3}

For the basic $D^{(1)}_n$-automaton, 
the time evolution $\mathcal{T}_1$ in 
Definition \ref{def:calT} 
was described as a motion of particles and anti-particles 
that undergo pair creation and annihilation \cite{HKT3}.
Here we generalize the result to the inhomogeneous 
$D^{(1)}_n$-automaton for any $\mathcal{T}_a$.
It turns out that the inhomogeneous automaton 
admits a reduction to the basic one 
except inserting certain operator $Q$ which 
will be introduced around (\ref{eq:Qdef}).

First we prepare some notations for the basic automaton.
We identify $B_1$ with $J$ as a set via the correspondence 
$\vec a \in B_1 \leftrightarrow a \in J$.
Then states of the basic $D^{(1)}_n$-automaton 
are infinite arrays of the letters {}from $J$.
The totality of them will be denoted by
\begin{equation}\label{eq:tildeW}
\widetilde{W}[a] = \{ (\ldots, b_i, b_{i+1},\ldots) \in
\cdots \times B_1 \times B_1 \times \cdots \mid  b_i=a
\text{ for } \vert i \vert \gg 1 \}.
\end{equation}

In the sequel, we introduce the operators $P, Q, L_j$ and $K_j$.
Except $P$, they are actually dependent on the choice $a$ of the 
boundary condition.
(In appendix \ref{app:b} we will write them as $Q^{(a)}$ and $K^{(a)}_j$.)
For
$\vec{x}=(x_1,\ldots,x_n,x_{-n},\ldots,x_{-1}) \in B_k$, 
let $P(\vec{x})$ be an array of $k$ letters arranged as
\begin{equation}\label{eq:pdef1}
P(\vec{x}) = \overbrace{-1 \ldots -\!1}^{x_{-1}}\ldots
\overbrace{-n \ldots -\!n}^{x_{-n}}
\overbrace{n \ldots n \vphantom{1}}^{x_n}
\ldots
\overbrace{1 \ldots 1}^{x_1}.
\end{equation}
For $\boldsymbol{b}=(\ldots, \vec{b}_i, \vec{b}_{i+1},\ldots) \in
W[a]_l$,
let $P(\boldsymbol{b})$ be an array of infinite letters
and walls given by
\begin{equation}\label{eq:pb}
P(\boldsymbol{b}) = \ldots \vert P(\vec{b}_i) \vert
P(\vec{b}_{i+1}) \vert \ldots ,
\end{equation}
where each $P(\vec{b}_i)$ for $\vec{b}_i \in B_{l_i}$
is defined by (\ref{eq:pdef1}).
The symbol $\vert$ denotes the wall inserted between
every $P(\vec{b}_i)$ and $P(\vec{b}_{i+1})$.

Let $\widetilde{W}[a]_l$ be the same set as 
$\widetilde{W}[a]$ except that walls are inserted at the same positions
as in the elements of $P(W[a]_l)$.
Namely $\widetilde{W}[a]_l$ consists of the elements of the form
\begin{displaymath}
\boldsymbol{c} = \ldots \vert C_i \vert C_{i+1} \vert \ldots ,
\end{displaymath}
where $C_i (i \in \Z)$, which we call a {\em cell}, 
is an array of $l_i$ letters {}from $J$.
For distant $i$, i.e., $\vert i \vert \gg 1$, 
the cell $C_i$ consists only of the
letter $a$. On the other hand $C_i$ with not necessarily distant $i$
can assume different ordering {}from (\ref{eq:pdef1}), hence 
$P(W[a]_l) \subset \widetilde{W}[a]_l$ holds.

We let $\widetilde{W}[a]^Q_l$
denote a subset of $\widetilde{W}[a]_l$ in which every 
cell $C_i$  has the form
\begin{equation}
\label{eq:cell}
C_i=
\overbrace{-a \ldots -\!a }^{s_i}
\ldots
\ldots
\overbrace{a \ldots a}^{t_i},
\end{equation}
involving $\pm a$ nowhere else in the cell $C_i$.
We let $Q$ be the operator on
$\widetilde{W}[a]^Q_l$
that sends each cell $C_i$ in (\ref{eq:cell}) to
\begin{equation}\label{eq:Qdef}
C'_i=
\overbrace{a \ldots a}^{t_i}
\ldots
\ldots
\overbrace{-a \ldots -\!a }^{s_i},
\end{equation}
and keeps the other letters unchanged.
By the definition $Q$ becomes trivial
for the basic automaton.

For $a \in J$ and $j \in J \backslash \{a, -a\}$
we introduce a map  $L_j: (\Z_{\ge 0}) \times B_1
\rightarrow B_1 \times (\Z_{\ge 0})$ as
follows.
Let the diagram
\begin{equation}\label{eq:batsu}
\batten{m}{b}{b'}{m'}{j}
\end{equation}
depict the relation $L_j:(m,b) \mapsto (b',m')$.
Then under the assumption 
$m \in \mathbb{Z}_{\ge 0}$ and $b \in B_1
\backslash \{ j, -j, a, -a \}$, the following specify 
the map $L_j$ completely:
\\
\begin{equation}\label{eq:batt}
\batten{m\!+\!1}{-j}{-a}{m}{j}\hspace{3pt}
\batten{0}{-j}{-j}{0}{j}\hspace{3pt}
\batten{m}{-a}{-j}{m\!+\!1}{j}\hspace{21pt}
\batten{m}{j}{a}{m\!+\!1}{j}\hspace{33pt}
\batten{m\!+\!1}{a}{j}{m}{j}\hspace{3pt}
\batten{0}{a}{a}{0}{j}\hspace{1pt}
\batten{m}{b}{b}{m}{j}
\end{equation}
\par\noindent
For $j \in J \backslash \{a, -a\}$
we construct an operator
$K_j: \widetilde{W}[a] \rightarrow \widetilde{W}[a]$ 
{}from a composition of $L_j$'s as follows.
Given a state of the basic automaton $(\ldots, b_i, b_{i+1}, \ldots) \in
\widetilde{W}[a]$,
there exists an integer $g$ such that $b_{g'} = a$ for all $g' < g$
owing to the boundary condition  (\ref{eq:tildeW}).
Fix any such $g$.
Then the operator $K_j$ maps
$(\ldots, b_i, b_{i+1}, \ldots)$ to $(\ldots, c_i, c_{i+1}, \ldots)$,
where $c_{g'} = b_{g'}$ for all $g' < g$.
The remaining $c_{g}, c_{g+1},\ldots$ are determined by the composition of
$L_j$'s  as

\begin{center}
\begin{equation}\label{eq:LL}
\begin{picture}(80,20)(-12,0)
	\put(0,0){\vector(1,0){70}}
	\multiput(10,0)(20,0){3}{
		{\thicklines\put(0,10){\vector(0,-1){20}}}
	}
	\put(10,12){\makebox(0,0)[b]{$b_g$}}
	\put(30,12){\makebox(0,0)[b]{$b_{g+1}$}}
	\put(50,12){\makebox(0,0)[b]{$\cdots$}}
	\put(10,-12){\makebox(0,0)[t]{$c_g$}}
	\put(30,-12){\makebox(0,0)[t]{$c_{g+1}$}}
	\put(50,-12){\makebox(0,0)[t]{$\cdots$}}
	\put(-2,0){\makebox(0,0)[r]{0}}
\end{picture}
\end{equation}
\end{center}
\vskip25pt
It is easy to see that the result is independent of the choice of $g$.
We let $K_j$ act also on $\widetilde{W}[a]_l$ 
in the same way as $\widetilde{W}[a]$ by ignoring 
the walls.
For an example if $a=3$, 
the action of $K_2$ on a state 
$\ldots, -2, 2, -3, 3, 4, -3, -2, 3, 3,\ldots \in \widetilde{W}[3]$ proceeds as
($\ldots$ denotes the array of $3$ only) 

\vspace{0.3cm}
\setlength{\unitlength}{10pt}
\begin{tabular}{rp{1pt}cp{1pt}cp{1pt}cp{1pt}cp{1pt}cp{1pt}cp{1pt}cp{1pt}cp{1pt}cp{1pt}cp{1pt}}
&&   -2& &   2& &    -3& &    3& &   4& &    -3& &    -2& &    3& &    3& \\
\yajirusiK{2}
&
0&\batu{-2}&0&\batu{-2}&1&\batu{-2}&2&\batu{-2}&1&
\batu{-2}&1&\batu{-2}&2&\batu{-2}&1&\batu{-2}&0&\batu{-2}&0\\
&&    -2& &   3& &  -2& &  2& & 4& &  -2& &   -3& &    2& &    3& \\
\end{tabular}\\
\vspace{0.2cm}

\noindent
leading to the output $\ldots, -2, 3, -2, 2, 4, -2, -3, 2, 3, \ldots
\in \widetilde{W}[3]$.

Let us present an interpretation as a system of particles and 
anti-particles with various colors.
A state $(\ldots, b_i, b_{i+1},\ldots) \in \widetilde{W}[a]$ 
is regarded as
a one dimensional array of boxes.
The local state $b_i=b$ with $b \in 
B_1 \setminus \{a, -a\}$ means that the $i$-th box contains a particle of 
color $b$. 
Since the set $B_1 \setminus \{a, -a\}$ is invariant 
under the interchange $b \leftrightarrow -b$,
the local state $-b$ may be viewed as a box containing an anti-particle of 
$b$ and vice versa.
We regard the selected local state $a \in B_1$ 
as an empty box.
On the other hand, 
when acting $K_j$ on $\widetilde{W}[a]$,
the local state $-a \in B_1$ plays a special role as 
a bound state of a particle of color $j$ 
and its anti-particle $-j$.
The boundary condition 
in (\ref{eq:tildeW}) allows only finitely many boxes to be 
non-empty.
Under this interpretation, the operator $K_j$ 
describes the dynamics of right-moving particles of color $j$ 
seeking the empty box $a$ or a free partner, 
{\it i.e.}, an anti-particle $-j$ not yet paired with the other $j$'s, 
to form a bound state $-a$ within a box.
To see this, regard the diagram (\ref{eq:batsu}) 
as representing $m$ incident particles with color $j$ 
coming {}from the left and $m'$ particles with the same color $j$ 
outgoing to the right.
The propagation of the color $j$ particles to the right 
thereby induces the change of contents in a box 
corresponding to the transformation $b \rightarrow b'$.
In fact, in the first figure of (\ref{eq:batt}), 
a color $j$ particles coming {}from the 
left is trapped in a box to form the bound state $-a$ with 
the anti-particle $-j$ 
(pair annihilation $(j, -j) \rightarrow -a$).
In the second figure, there are no color $j$ particles 
coming {}from the left nor in the box hence nothing happens.
In the third figure, a color $j$ particle is 
removed off the bound state $-a$ leaving the 
anti-particle $-j$ 
(pair creation $-a \rightarrow (j, -j)$).
In the fourth figure,
a color $j$ particle is taken away to the right 
leaving an empty box $a$.
In the fifth figure, a color $j$ particle stops 
at the empty box to fill it.
In the last two figures, nothing can happen.
When $K_j$ acts on $\widetilde{W}[a]_l$ 
instead of $\widetilde{W}[a]$, it induces the same motion of 
particles without any effect {}from the walls.
In contrast with this, the operator $Q$
remembers the walls at $l = (l_i)_{i \in \Z}$ and 
swaps the positions of the empty boxes $a$ and the bound states 
$-a$ within each cell.

\begin{definition}\label{def:T}
For $1 \leq a \leq n$ we define
the operator $T_a$ on $W[a]_l$
as
\begin{align}
T_a &= P^{-1} \left( \prod_{j=a+1}^{n} K_j \right)
\left( \prod_{j=-n}^{-(a+1)} K_j \right)
\left( \prod_{j=-(a-1)}^{-1} K_j \right)
Q \left( \prod_{j=1}^{a-1} K_j \right) P,
\label{eq:Tdef1}\\
\intertext{and $T_{-a}$ on $W[-a]_l$ as}
T_{-a} &= P^{-1} \left( \prod_{j=-(a-1)}^{-1} K_j \right)
Q
\left( \prod_{j=1}^{a-1} K_j \right)
\left( \prod_{j=a+1}^{n} K_j \right)
\left( \prod_{j=-n}^{-(a+1)} K_j \right) P,
\label{eq:Tdef2}
\end{align}
where $\prod_{j=c}^{c-1}K_j = id$, 
and for $c \le d$ we adopt the convention 
$\prod_{j=c}^{d}K_j = K_cK_{c+1}\cdots K_d$.
\end{definition}
One should remember that the operators $K_j$ and $Q$  (and $L_j$) 
are actually dependent on $a$ as opposed to $P$.
Up to this distinction,
the expressions of $T_n$ and $T_{-n}$ look formally identical.
See Remark \ref{rem:KK} for the possible rearrangement of these 
products.

\begin{example}\label{ex:Td4}
The time evolution operators for inhomogeneous $D^{(1)}_4$-automaton 
read as 
\begin{align*}
&T_1 = P^{-1}K_2K_3K_4K_{-4}K_{-3}K_{-2}QP,\\
&T_2 = P^{-1}K_3K_4K_{-4}K_{-3}K_{-1}QK_1P,\\
&T_3 = P^{-1}K_4K_{-4}K_{-2}K_{-1}QK_1K_2P,\\
&T_4 = P^{-1}K_{-3}K_{-2}K_{-1}QK_1K_2K_3P,\\
&T_{-4} = P^{-1}K_{-3}K_{-2}K_{-1}QK_1K_2K_3P,\\
&T_{-3} = P^{-1}K_{-2}K_{-1}QK_1K_2K_4K_{-4}P,\\
&T_{-2} = P^{-1}K_{-1}QK_1K_3K_4K_{-4}K_{-3}P,\\
&T_{-1} = P^{-1}QK_2K_3K_4K_{-4}K_{-3}K_{-2}P.
\end{align*}
\end{example}
Each operator $K_j\, (j \in J)$ in Definition \ref{def:T} 
transforms the cells as
\begin{align}
&\ldots \overbrace{a ... a}^\alpha 
\overbrace{j ... j}^\beta \ldots
\overbrace{-\!j ... -\!j}^\gamma 
\overbrace{-a ... -\!a}^\delta \ldots 
\mapsto
\ldots \overbrace{j ... j}^{\alpha'}
\overbrace{a ... a}^{\beta'} \ldots
\overbrace{-a ... -\!a}^{\gamma'}
\overbrace{-\!j ... -\!j}^{\delta'} \ldots,\label{eq:ord1}
\intertext{for $-n \le j \le -1$, and  as }
&\ldots \overbrace{-\!j ... -\!j}^\gamma 
\overbrace{-a ... -\!a}^\delta \ldots 
\overbrace{a ... a}^\alpha 
\overbrace{j ... j}^\beta \ldots
\mapsto
\ldots
\overbrace{-a ... -\!a}^{\gamma'}
\overbrace{-\!j ... -\!j}^{\delta'} \ldots
\overbrace{j ... j}^{\alpha'}
\overbrace{a ... a}^{\beta'} \ldots,\label{eq:ord2}
\end{align}
for $1 \le j \le n$, where 
$\alpha + \beta = \alpha'+\beta'$ and 
$\gamma + \delta = \gamma' + \delta'$.
The letters other than $\pm j$ and $\pm a$ are kept unchanged.
In (\ref{eq:ord1}) and (\ref{eq:ord2}), 
the ordering of the letters are the same as 
(\ref{eq:pdef1}) except the positions of $\pm a$.
See Example \ref{ex:t3}.
This fact assures that the right hand sides of
(\ref{eq:Tdef1}) and (\ref{eq:Tdef2}) without the leftmost $P^{-1}$
belong to $P(W[a]_l)$, hence 
Definition \ref{def:T} makes sense.
\begin{example}\label{ex:kactionD}
Consider $T_{-4}$ for $D^{(1)}_4$ given in Example \ref{ex:Td4}.
The operators $K_j$ and $Q$ therein act as 
(to save the space we write $\bar 4$ to mean $-4$ etc.) 
\begin{align*}
&\ldots \vert \bar 3 4 3 1 \vert \bar 1 \bar 2 \vert \bar 4 2 
\vert \bar 4 3 1 \vert \bar 4 \bar 4 \bar 4 \vert \ldots \\
\overset{K_3}{\longmapsto} \quad
&\ldots \vert \bar 3 \bar 3 \bar 4 1 \vert \bar 1 \bar 2 \vert 3 2 
\vert 3 \bar 4 1 \vert 3 \bar 4 \bar 4 \vert \ldots \\
\overset{K_2}{\longmapsto} \quad
&\ldots \vert \bar 3 \bar 3 \bar 4 1 \vert \bar 1 \bar 2 \vert 3 \bar 4
\vert 3 2 1 \vert 3 \bar 4 \bar 4 \vert \ldots \\
\overset{K_1}{\longmapsto} \quad
&\ldots \vert \bar 3 \bar 3 \bar 4 \bar 4 \vert 4 \bar 2 \vert 3 \bar 4
\vert 3 2 \bar 4 \vert 3 1 \bar 4 \vert \ldots \\
\overset{Q}{\longmapsto} \quad
&\ldots \vert \bar 4 \bar 4 \bar 3 \bar 3 \vert  \bar 2 4 \vert  \bar 4 3
\vert \bar 4 3 2 \vert \bar 4 3 1 \vert \ldots \\
\overset{K_{-1}}{\longmapsto} \quad
&\ldots \vert \bar 4 \bar 4 \bar 3 \bar 3 \vert  \bar 2 1 \vert  \bar 1 3
\vert \bar 4 3 2 \vert \bar 4 3 1 \vert \ldots \\
\overset{K_{-2}}{\longmapsto} \quad
&\ldots \vert \bar 4 \bar 4 \bar 3 \bar 3 \vert  \bar 4 1 \vert  \bar 1 3
\vert \bar 2 3 2 \vert \bar 4 3 1 \vert \ldots \\
\overset{K_{-3}}{\longmapsto} \quad
&\ldots \vert \bar 4 \bar 4 \bar 4 \bar 4 \vert  \bar 3 1 \vert  \bar 1 4
\vert \bar 2 3 2 \vert \bar 4 3 1 \vert \ldots,
\end{align*}
where $\ldots$ stands for the cells that consist of $\bar 4$ only.
Both the initial and the final states belong to 
$P(W[-4]_{l})$ with $l=(\ldots,4,2,2,3,3,\ldots)$.
\end{example}
For $a \neq \pm n$, the time evolution $T_a$ involves the factor 
$K_nK_{-n}$. At the intermediate state between $K_n$ and $K_{-n}$, 
the cells can contain the letters $n$ and $-n$ simultaneously, but 
$a$ and $-a$ can not coexist.
For the other intermediate states, the situation is opposite, namely,
$a$ and $-a$ can coexist but $n$ and $-n$ can not.
The following is such an example. See also Example \ref{ex:t3}.
\begin{example}\label{ex:kactionD2}
Consider $T_{1}$ for $D^{(1)}_4$ given in Example \ref{ex:Td4}.
The operators $K_j$ and $Q$ therein act as 
(to save the space we write $\bar 4$ to mean $-4$ etc.) 
\begin{align*}
&\ldots \vert \bar 4 \bar4 \vert \bar3 4 3 1
\vert \bar1 \bar3 \bar4 \vert \bar2 \vert \bar1 1 \vert 1 1 1 \vert \ldots \\
\overset{Q}{\longmapsto} \quad
&\ldots \vert \bar 4 \bar4 \vert 1 \bar3 4 3 
\vert \bar3 \bar4 \bar1 \vert \bar2 \vert 1 \bar1 \vert 1 1 1 \vert \ldots \\
\overset{K_{-2}}{\longmapsto} \quad
&\ldots \vert \bar 4 \bar4 \vert 1 \bar3 4 3 
\vert \bar3 \bar4 2 \vert 1 \vert \bar2 2 \vert \bar2 \bar2 1 \vert \ldots \\
\overset{K_{-3}}{\longmapsto} \quad
&\ldots \vert \bar 4 \bar4 \vert 1 1 4 \bar1
\vert 1 \bar4 2 \vert \bar3 \vert \bar2 2 \vert \bar2 \bar2 1 \vert \ldots \\
\overset{K_{-4}}{\longmapsto} \quad
&\ldots \vert 1 1 \vert \bar4 \bar4 4 4 
\vert \bar4 1 2 \vert \bar3 \vert \bar2 2 \vert \bar2 \bar2 \bar4 \vert \ldots \\
\overset{K_{4}}{\longmapsto} \quad
&\ldots \vert 1 1 \vert \bar4 \bar4 1 1  
\vert \bar1 4 2 \vert \bar3 \vert \bar2 2 \vert \bar2 \bar2 \bar4 \vert \ldots \\
\overset{K_{3}}{\longmapsto} \quad
&\ldots \vert 1 1 \vert \bar4 \bar4 1 1  
\vert \bar3 4 2 \vert \bar1 \vert \bar2 2 \vert \bar2 \bar2 \bar4 \vert \ldots \\
\overset{K_{2}}{\longmapsto} \quad
&\ldots \vert 1 1 \vert \bar4 \bar4 1 1  
\vert \bar3 4 1 \vert \bar2 \vert \bar1 1 \vert \bar1 \bar1 \bar4 \vert \ldots,
\end{align*}
where $\ldots$ stands for the cells that consist of $1$ only.
Both the initial and the final states belong to 
$P(W[1]_{l})$ with $l=(\ldots,2,4,3,1,2,3,\ldots)$.
\end{example}
Now we give the main theorem of this paper.
\begin{theorem}\label{th:main}
For any $a \in J$, $\mathcal{T}_a = T_a$ is valid.
Namely, the time evolution operator $\mathcal{T}_a$ in
Definition \ref{def:calT} coincides with 
$T_a$ in Definition \ref{def:T} that admits the particle interpretation.
\end{theorem}

We include a proof in section \ref{sec:5} and appendix \ref{app:b}.

\section{\mathversion{bold}
Embedding into inhomogeneous $D^{(1)}_n$-automaton}\label{sec:4}

In this section an automaton means an inhomogeneous automaton.
The $D^{(1)}_n$-automaton ($n \ge 4$) contains, as certain sectors, 
all the automata associated with the 
Kang-Kashiwara-Misra (KKM) crystals \cite{KKM} for the 
other non-exceptional affine Lie algebras 
$A^{(2)}_{2n-3}$, 
$A^{(2)}_{2n-4}$, 
$B^{(1)}_{n-1}$, 
$C^{(1)}_{n-2}$ and 
$D^{(2)}_{n-1}$.
Here we describe the embedding of these automata 
along the following scheme:

\vspace{0.6cm}

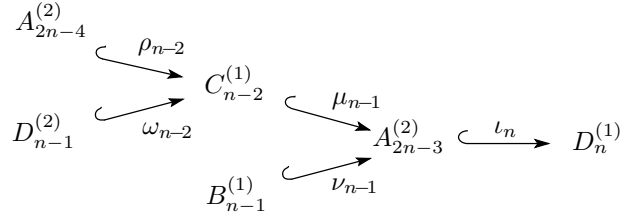
\unitlength 0.1in
\begin{picture}( 33.2300,  9.1700)(  6.000,-14.7200)
\put(17.0000,-6.4000){\makebox(0,0){$A^{(2)}_{2n-4}$}}%
%
\special{pn 8}%
\special{pa 2960 1470}%
\special{pa 3370 1362}%
\special{fp}%
\special{sh 1}%
\special{pa 3370 1362}%
\special{pa 3300 1360}%
\special{pa 3318 1376}%
\special{pa 3312 1398}%
\special{pa 3370 1362}%
\special{fp}%
%
\special{pn 8}%
\special{pa 2960 1470}%
\special{pa 2928 1472}%
\special{pa 2908 1452}%
\special{pa 2914 1422}%
\special{pa 2940 1406}%
\special{pa 2944 1404}%
\special{sp}%
%
\special{pn 8}%
\special{pa 1980 1160}%
\special{pa 2390 1052}%
\special{fp}%
\special{sh 1}%
\special{pa 2390 1052}%
\special{pa 2320 1050}%
\special{pa 2338 1066}%
\special{pa 2332 1088}%
\special{pa 2390 1052}%
\special{fp}%
%
\special{pn 8}%
\special{pa 1980 1160}%
\special{pa 1948 1162}%
\special{pa 1928 1142}%
\special{pa 1934 1112}%
\special{pa 1960 1096}%
\special{pa 1964 1094}%
\special{sp}%
%
\special{pn 8}%
\special{pa 2960 1100}%
\special{pa 3368 1212}%
\special{fp}%
\special{sh 1}%
\special{pa 3368 1212}%
\special{pa 3310 1174}%
\special{pa 3318 1198}%
\special{pa 3298 1214}%
\special{pa 3368 1212}%
\special{fp}%
%
\special{pn 8}%
\special{pa 2960 1100}%
\special{pa 2932 1086}%
\special{pa 2922 1058}%
\special{pa 2944 1036}%
\special{pa 2976 1034}%
\special{pa 2978 1036}%
\special{sp}%
%
\special{pn 8}%
\special{pa 1970 840}%
\special{pa 2384 934}%
\special{fp}%
\special{sh 1}%
\special{pa 2384 934}%
\special{pa 2322 900}%
\special{pa 2332 922}%
\special{pa 2314 940}%
\special{pa 2384 934}%
\special{fp}%
%
\special{pn 8}%
\special{pa 1970 840}%
\special{pa 1942 828}%
\special{pa 1932 798}%
\special{pa 1952 778}%
\special{pa 1982 774}%
\special{pa 1986 774}%
\special{sp}%
\put(45.6000,-12.5000){\makebox(0,0){$D^{(1)}_n$}}%
\put(26.6000,-9.7000){\makebox(0,0){$C^{(1)}_{n-2}$}}%
\put(26.7000,-15.5000){\makebox(0,0){$B^{(1)}_{n-1}$}}%
%
\special{pn 8}%
\special{pa 3880 1284}%
\special{pa 4304 1284}%
\special{fp}%
\special{sh 1}%
\special{pa 4304 1284}%
\special{pa 4236 1264}%
\special{pa 4250 1284}%
\special{pa 4236 1304}%
\special{pa 4304 1284}%
\special{fp}%
%
\special{pn 8}%
\special{ar 3880 1250 48 34  1.5707963 4.7123890}%
\put(16.6000,-12.2000){\makebox(0,0){$D^{(2)}_{n-1}$}}%
\put(35.7000,-12.5000){\makebox(0,0){$A^{(2)}_{2n-3}$}}%
\put(22.8000,-7.8000){\makebox(0,0){$\rho_{n\!-\!2}$}}%
\put(23.1000,-12.2000){\makebox(0,0){$\omega_{n\!-\!2}$}}%
\put(32.9000,-15.1000){\makebox(0,0){$\nu_{n\!-\!1}$}}%
\put(33.0000,-10.6000){\makebox(0,0){$\mu_{n\!-\!1}$}}%
\put(40.8000,-12.1000){\makebox(0,0){$\iota_n$}}%
\end{picture}%
\setlength{\unitlength}{0.35mm}

\vspace{0.8cm}\noindent
By embedding of an automaton, we precisely mean the following.
Take any pair $\psi: X \hookrightarrow X'$ in the above scheme.
Then there is an injection 
$\psi: B_k \rightarrow B'_{k'} $ 
between the KKM crystals associated with these algebras 
$X$ and $X'$, respectively.
Moreover, this $\psi$ has an extension 
to Kashiwara operators \cite{K} 
$e_i, f_i$ 
in such a way that $\psi(e_i \vec x) = \psi(e_i)\psi(\vec x)$ and 
$\psi(f_i \vec x) = \psi(f_i)\psi(\vec x)$ for any 
$\vec x \in B_{k}$ and $i$. 
It follows that the combinatorial $R$ for $X$ 
is embedded into the one for $X'$ as 
$R = (\psi^{-1}\ot \psi^{-1})R'(\psi \ot \psi)$. 
The map $\psi^\Delta = 
(\cdots \otimes \psi \otimes \psi \otimes \cdots)$ 
embedds the states of the $X$-automaton into $X'$-automaton
and relates their time evolutions simply by 
$T^X_a  = (\psi^\Delta)^{-1}T^{X'}_{a'}\psi^\Delta$.
The precise correspondence of the boundary conditions $a$ and $a'$
will be described below together with 
the injections $\iota_n, \mu_n, \nu_n, \rho_n$ and $\omega_n$.
The embedding $\rho_n$ and $\omega_n$ for the basic automata 
have appeared in \cite{HKOTY}.
The $\mu_n$ and $\iota_n$ have been introduced in 
section 6.1 of \cite{KOTY2} in a tropical setting. 

In Table \ref{tab:B}, the KKM crystal $B_k$ is listed 
for each algebra as a set.
They all have the presentation 
$B_k = \{ \vec x=(x_1,\ldots, x_{-1}) \in \Z_{\ge 0}^N 
\mid (\ast) \}$ with certain $N$ and the condition $(\ast)$ 
on $\vec x$.
We let $s(\vec x) = x_1 + \cdots + x_{-1}$ 
denote the sum of all the components of $\vec x$ and list the 
array $\vec x$ and $(\ast)$.
\begin{table}\caption{Sets $B_k=\{\vec x \mid (\ast)\}$}\label{tab:B}
\begin{center}
\begin{tabular}[h]{c|c|c}
algebra& array $\vec x$ & condition $(\ast)$ \\\hline 
%
$A^{(2)}_{2n-1}$
&$(x_1,\ldots, x_{n},x_{-n},\ldots, x_{-1})\in \Z_{\ge 0}^{2n}$
&$s(\vec x)=k$\\[5pt]
$A^{(2)}_{2n}$
&$(x_1,\ldots, x_{n},x_{-n},\ldots, x_{-1})\in \Z_{\ge 0}^{2n}$
&$s(\vec x)\le k$\\[5pt]
$B^{(1)}_n$
&$(x_1,\ldots, x_{n},x_0,x_{-n},\ldots, x_{-1})\in \Z_{\ge 0}^{2n+1}$
&$x_0 \in \{0,1\},\; s(\vec x)=k$\\[5pt]
$C^{(1)}_n$
&$(x_1,\ldots, x_{n},x_{-n},\ldots, x_{-1})\in \Z_{\ge 0}^{2n}$
&$s(\vec x) \in k-2\Z_{\ge 0}$\\[5pt]
$D^{(1)}_n$
&$(x_1,\ldots, x_{n},x_{-n},\ldots, x_{-1})\in \Z_{\ge 0}^{2n}$
&$x_nx_{-n}=0,\; s(\vec x)=k$\\[5pt]
$D^{(2)}_{n+1}$
&$(x_1,\ldots, x_{n},x_0,x_{-n},\ldots, x_{-1})\in \Z_{\ge 0}^{2n+1}$
&$x_0 \in \{0,1\},\; s(\vec x)\le k$\\[5pt]
\end{tabular}
\end{center}
\end{table}
The images of the maps 
appearing in the embedding scheme are given in 
Table \ref{tab:alpha}.
\begin{table}\caption{Maps $\iota_n, \mu_n, \nu_n, \rho_n, \omega_n$}
\label{tab:alpha}
\begin{center}
\begin{tabular}[h]{c|c|c}
$\psi$ & $\psi(\vec x)$ & $\psi(e_i)$ \\\hline 
$\iota_{n}: B_k\rightarrow B'_k$
&$(x_1,\ldots, x_{n-1},0,0,x_{-n+1},\ldots, x_{-1})$
&$e_i$\\[5pt]
$\mu_n: B_k\rightarrow B'_k$
&$(x_0,x_1,\ldots, x_{n-1},x_{-n+1},\ldots, x_{-1},x_0)$
&$e_{i+1}$\\[5pt]
$\nu_n: B_k\rightarrow B'_{2k}$
&$(2x_1,\ldots, 2x_{n-1},2x_n+x_0,2x_{-n}+x_0,
2x_{-n+1},\ldots, 2x_{-1})$
&$e_{i}^{2-\delta_{i,n}}$\\[5pt]
$\rho_n: B_k\rightarrow B'_{2k}$
&$(2x_1,\ldots, 2x_{n},2x_{-n},\ldots, 2x_{-1})$
&$e_{i}^{2-\delta_{i,0}}$\\[5pt]
$\omega_n: B_k\rightarrow B'_{2k}$
&$(2x_1,\ldots, 2x_{n-1},2x_n+x_0,2x_{-n}+x_0,
2x_{-n+1},\ldots, 2x_{-1})$
&$e_{i}^{2-\delta_{i,0}-\delta_{i,n}}$\\[5pt]
\end{tabular}
\end{center}
\end{table}
The component $x_0$ appearing in the image $\mu_n(\vec x)$ is 
the non-negative integer specified by  
$x_0 = (k-\sum_{j=1}^{n-1}(x_j+x_{-j}))/2$.
The image $\psi(f_i)$ is obtained {}from $\psi(e_i)$ 
by changing the letter $e$ to $f$.

As an example let us illustrate the embedding $\iota_n$ in further detail. 
This is a significant case in that
the $A^{(2)}_{2n-3}$-automaton is considerably simpler than the 
$D^{(1)}_n$-automaton 
and all the other automata can firstly be embedded into the former.

In view of $\iota_n: (x_1,\ldots,x_{n-1},x_{-n+1},\ldots,x_{-1})
\mapsto (x_1,\ldots,x_{n-1},0,0,x_{-n+1},\ldots,x_{-1})$, 
the $A^{(2)}_{2n-3}$-automaton is a restriction of the 
$D^{(1)}_n$-automaton to those local states 
satisfying $x_n = x_{-n}=0$.
The set of states is formally the same as
(\ref{eq:W}) once $B_{l_i}$ is understood as the one for $A^{(2)}_{2n-3}$.
The boundary condition is 
specified by a letter $a \in \{\pm1, \ldots, \pm(n-1)\}$.
The time evolutions $T_a$ for these $a$ are 
still given by the formulas in Definition \ref{def:T} 
except only that the states they act are restricted.
The $T_a$ with $a \neq \pm n$ contains 
the operators $K_n$ and $K_{-n}$ only through the 
combination $K_nK_{-n}$. See Example \ref{ex:Td4} for $n=4$.
In Lemma \ref{lem:k-a} we will show that the composition 
$K_nK_{-n}$ 
under the restriction $x_n = x_{-n}=0$ coincides with 
a simple operator $K_{-a}$ defined as follows.
(Note that $K_{-a}$ was not introduced in $D^{(1)}_n$ case.)
The $K_{-a}$ is the composition of 
$L_{-a}$ in the same was as (\ref{eq:LL}), and the $L_{-a}$ is 
specified by the diagrams:
\begin{equation}\label{eq:batt2}
\batten{m}{-a}{a}{m\!+\!1}{-a}\hspace{34pt}
\batten{m\!+\!1}{a}{-a}{m}{-a}\hspace{5pt}
\batten{0}{a}{a}{0}{-a}\hspace{6pt}
\batten{m}{b}{b}{m}{-a}
\end{equation}
\par\noindent
where $b \in B\setminus \{a,-a\}$ and $m \in \Z_{\ge 0}$.
These diagrams represent the motion of the right-moving 
particles $-a$ seeking the empty box $a$ without 
pair creation nor annihilation.
Such a behavior of color $-a$ particles is exceptional, in that 
they still act as bound states of $j$ and $-j$ under the 
operator $K_j$ with $j \neq -a$.
\begin{example}\label{ex:Ta2}
The time evolution operators for $A^{(2)}_5$-automaton read as
\begin{align*}
&T_1 = P^{-1}K_2K_3K_{-1}K_{-3}K_{-2}QP,\\
&T_2 = P^{-1}K_3K_{-2}K_{-3}K_{-1}QK_{1}P,\\
&T_3 = P^{-1}K_{-3}K_{-2}K_{-1}QK_{1}K_{2}P,\\
&T_{-3} = P^{-1}K_{-2}K_{-1}QK_{1}K_{2}K_{3}P,\\
&T_{-2} = P^{-1}K_{-1}QK_{1}K_{3}K_{2}K_{-3}P,\\
&T_{-1} = P^{-1}QK_{2}K_{3}K_{1}K_{-3}K_{-2}P.
\end{align*}
\end{example}
\begin{example}\label{ex:kactionA2}
Consider $T_{3}$ for $A^{(2)}_5$ given in Example \ref{ex:Ta2}.
The operators $K_j$ and $Q$ therein act as 
(to save the space we write $\bar 3$ to mean $-3$ etc.) 
\begin{align*}
&\ldots \vert 1 1 \vert \bar 1 \bar 1 \bar 3 1 
\vert \bar 1 3 2 \vert 3 3 3 3 \vert \ldots \\
\overset{K_2}{\longmapsto} \quad
&\ldots \vert 1 1 \vert \bar 1 \bar 1 \bar 2 1 
\vert \bar 1 2 3 \vert 2 3 3 3 \vert \ldots \\
\overset{K_1}{\longmapsto} \quad
&\ldots \vert 3 3 \vert \bar 3 \bar 3 \bar 2 3 
\vert \bar 3 2 3 \vert 2 3 3 3 \vert \ldots \\
\overset{Q}{\longmapsto} \quad
&\ldots \vert 3 3 \vert 3 \bar 2 \bar 3 \bar 3 
\vert 3 2 \bar 3 \vert 3 3 3 2 \vert \ldots \\
\overset{K_{-1}}{\longmapsto} \quad
&\ldots \vert 3 3 \vert 3 \bar 2 1 1
\vert \bar 1 2 1 \vert \bar 1 \bar 1 3 2 \vert \ldots \\
\overset{K_{-2}}{\longmapsto} \quad
&\ldots \vert 3 3 \vert 3 3 1 1
\vert \bar 1 \bar 3 1 \vert \bar 1 \bar 1 3 2 \vert \ldots \\
\overset{K_{-3}}{\longmapsto} \quad
&\ldots \vert 3 3 \vert 3 3 1 1
\vert \bar 1 3 1 \vert \bar 1 \bar 1 \bar 3 2 \vert \ldots,
\end{align*}
where $\ldots$ stands for the cells that consist of $3$ only.
\end{example}

Let us briefly comment on the $A^{(1)}_{n-1}$ case where the 
associated automaton is known as the (generalized) box-ball system
\cite{TS,T,HHIKTT,FOY}.
The relevant crystal $B_k$ is the set 
$\{\vec x = (x_1, \ldots, x_n) \in \Z^n_{\ge 0} 
\mid x_1 + \cdots + x_n = k \}$ equipped with the 
functions $\varphi_i(\vec x) = x_i$, 
$\varepsilon_i(\vec x) = x_{i+1}$ and the Kashiwara operators 
$e_i(\vec x) = (\ldots, x_i+1,x_{i+1}-1,\ldots)$, 
$f_i(\vec x) = (\ldots, x_i-1,x_{i+1}+1,\ldots)$, where all the 
indices are considered to be in $\Z/n\Z$.
We set $J = \{1,2,\ldots, n\}$ and 
$\vec a = (x_i = k\delta_{i,a}) \in B_k$ for $a \in J$.
In particular $\vec a \in B_1$ will simply be called $a$.
Under these notations the automaton states 
are again specified by (\ref{eq:W}).
Due to the Dynkin diagram $\Z_n$ symmetry, 
there is no loss of generality to restrict ourselves to the case 
$a=1$.
Then the time evolution is given by the Weyl group 
operators as ${\mathcal T}_1 = 
\sigma^\Delta S_2S_3\cdots S_{n-1}S_0$ \cite{HKT2},
where $\sigma^\Delta = \cdots \otimes \sigma \otimes \sigma \otimes \cdots$ 
is defined by $\sigma((x_1,x_2,\ldots,x_n)) = (x_2,\ldots, x_n,x_1)$.
In the same way as section \ref{sec:5} and appendix \ref{app:b},
we have proved that ${\mathcal T}_1 = T_1$ holds, where 
$T_1$ consists of the particle motion operator as
\begin{equation}\label{eq:agata1}
T_1 = P^{-1}K_2K_3\cdots K_nQP.
\end{equation}
Here $P$ acts on $W[1]_l$ as (\ref{eq:pb}) with (\ref{eq:pdef1}) 
simplified as 
\begin{equation}\label{eq:agata2}
P(\vec{x}) = \overbrace{n \ldots n \vphantom{1}}^{x_n}
\ldots
\overbrace{2 \ldots 2}^{x_2}\overbrace{1 \ldots 1}^{x_1}.
\end{equation}
The operator $Q$ rearranges each cell (\ref{eq:agata2}) into
$\overbrace{1 \ldots 1}^{x_1}\overbrace{n \ldots n \vphantom{1}}^{x_n}
\ldots
\overbrace{2 \ldots 2}^{x_2}$.
The operator $K_j$ is defined in terms of the vertex diagrams
($2 \le j \le n,\, b \neq 1,j$):
\begin{equation}\label{eq:batt3}
\batten{m}{j}{1}{m\!+\!1}{-a}\hspace{34pt}
\batten{m\!+\!1}{1}{j}{m}{-a}\hspace{5pt}
\batten{0}{1}{1}{0}{-a}\hspace{6pt}
\batten{m}{b}{b}{m}{-a}
\end{equation}
These rules coincide with $T_1$ of the $D^{(1)}_n$-automaton
restricted to the sector without anti-particles, i.e., 
those local states 
$(x_1,\ldots, x_n,x_{-n},\ldots,x_{-1})$ such that 
$x_{-n} = \cdots x_{-1}=0$.
In fact, the absence of anti-particles 
reduces the diagrams (\ref{eq:batt}) to the 
last four, which yield (\ref{eq:batt3}) 
under the choice $a=1$.
Thus we conclude that the inhomogeneous box-ball system is reduced to 
the basic case via (\ref{eq:agata1}).
Moreover, it coincides with a sector of the $D^{(1)}_n$-automaton 
without anti-particles.
The result equivalent to (\ref{eq:agata1}) has also been obtained 
in eq.(10) of \cite{HHIKTT} and 
by purely combinatorial arguments in \cite{F}.
The diagrams equivalent to 
(\ref{eq:batt3}) appeared in \cite{HIK} for the basic 
box-ball system.

\section{\mathversion{bold}
Weyl group operator $S_i$ as particle motion $K_j$}\label{sec:5}

Our main theorem 
$\mathcal{T}_a = T_a$ for $D^{(1)}_n$ is a 
composition of finer relations 
that intertwine
the Weyl group operators $S_i$ and 
the particle motion operators $K_j$.
The aim of this section is to present them 
in Proposition \ref{pr:main}.
Many assertions and 
formulas provided below for general case will be  
recognized most easily by 
consulting $n=4$ case in 
Examples \ref{ex:t1}, \ref{ex:t3} as well as 
Examples \ref{ex:calT} and \ref{ex:Td4}.

For
$\vec{x}=(x_1,\ldots,x_n,x_{-n},\ldots,x_{-1}) \in B_k$, 
let $\Pb(\vec{x})$ be an array of $k$ letters arranged as
\begin{equation}\label{eq:pbdef1}
\Pb(\vec{x}) = \overbrace{1 \ldots 1}^{x_1}
\overbrace{-2 \ldots -\!2}^{x_{-2}}\ldots
\overbrace{-n \ldots -\!n}^{x_{-n}}
\overbrace{n \ldots n \vphantom{1}}^{x_n}
\ldots
\overbrace{2 \ldots 2}^{x_2}
\overbrace{-1 \ldots -\!1}^{x_{-1}},
\end{equation}
which is obtained {}from $P(\vec x)$ in (\ref{eq:pdef1}) 
by interchanging the positions of $1$ and $-1$.
Similalry to (\ref{eq:pb}), we set 
$\Pb(\boldsymbol{b}) = \ldots \vert \Pb(\vec{b}_i) \vert
\Pb(\vec{b}_{i+1}) \vert \ldots$ for 
$\boldsymbol{b}=(\ldots, \vec{b}_i, \vec{b}_{i+1},\ldots) 
\in W[a]_l$.

In section \ref{sec:2} we introduced the Weyl group operator
$S_j$ acting on $\coprod_{a \in J}W[a]_l$. 
Here we use two further kinds of Weyl group operators
$r_j, s_j(0 \le j \le n)$.
The $r_j: B_1 \rightarrow B_1$ 
is the one for the crystal $B_1$.
In the crystal graph depicted in Figure \ref{fig:B},
if there is a color $j$ arrow coming into or outgoing 
from the node $p$, the image
$r_j(p)$ is determined by the relation 
$r_j(p) \overset{j}{\rightarrow} p$ or 
$r_j(p) \overset{j}{\leftarrow} p$, respectively.
If there is no such arrow, $r_j(p) = p$.
The $s_j$ is the one that acts on 
$\coprod_{a \in J}\widetilde{W}[a]_l$ totally componentwise
ignoring the walls. Namely, it transforms the cells as 
\begin{alignat*}{2}
\boldsymbol{c} &= \ldots \vert C_i \vert C_{i+1} \vert \ldots,
\qquad &
C_i &= p_1\ldots p_{l_i} \in B_1^{\times l_i},\\
s_j(\boldsymbol{c}) &= \ldots \vert C'_i \vert C'_{i+1} \vert \ldots,
\qquad &
C'_i &= r_j(p_1)\ldots r_j(p_{l_i}) \in B_1^{\times l_i}.
\end{alignat*}
Obviously, one has 
\begin{equation}
S_i(W[a]_l) \subseteq W[r_i(a)]_l,\qquad
s_i(\widetilde{W}[a]_l) \subseteq \widetilde{W}[r_i(a)]_l.
\label{eq:w1}
\end{equation}
For $a \in J$ let 
\begin{equation}\label{eq:calTexp}
\mathcal{T}_a = \sigma^\Delta S_{i_{2n-2}} \cdots S_{i_1}
\end{equation}
be the expression of $\mathcal{T}_a$ given in 
Definition \ref{def:calT}.
The sequence $i_1, i_2, \ldots, i_{2n-2}$ has the property
\begin{equation}\label{eq:iprop}
\varepsilon_{i_t}(r_{i_{t-1}}\cdots r_{i_2}r_{i_1}(a)) > 0,\quad 
1 \le t \le 2n-2,
\end{equation}
where $\varepsilon_i$ is defined in (\ref{eq:epdef}).
The operator $T_a$ in Definition \ref{def:T} 
is expressed as
\begin{equation}\label{eq:Texp}
\begin{split}
&T_a = P^{-1}Q^{\delta_{a,-1}}K'_{j_{2n-2}}\cdots 
K'_{j_1}P,\\
&K'_j \in \{K_j, K_jQ \},
\end{split}
\end{equation}
with the sequence $j_1, j_2, \ldots j_{2n-2} \in J$ specified by
\begin{equation}\label{eq:ij}
j_t = r_{i_1}r_{i_2}\cdots r_{i_{t-1}}r_{i_t}
r_{i_{t-1}}\cdots r_{i_2}r_{i_1}(a),\quad 1 \le t \le 2n-2.
\end{equation}
In view of (\ref{eq:w1}), 
the operator $S_{i_t}$ in (\ref{eq:calTexp}) 
acts among the sets:
\begin{equation}\label{eq:sww}
S_{i_t}: W[r_{i_{t-1}}\cdots r_{i_1}(a)]_l \rightarrow 
W[r_{i_{t}}\cdots r_{i_1}(a)]_l
\end{equation}
for $a \in J$ and $1 \le t \le 2n-2$.

Let $r: B_1 \rightarrow B_1$ be the involutive automorphism 
$1 \leftrightarrow -1$ and $n \leftrightarrow -n$ 
keeping the other letters unchanged.
It is easy to check that $r_{i_1}\cdots r_{i_{2n-2}}=r$ 
for all $a \in J$.
Consequently 
$s (=s^{-1}) = s_{i_1}\cdots s_{i_{2n-2}}: 
\coprod_{a \in J}\widetilde{W}[a]_l \rightarrow 
\coprod_{a \in J}\widetilde{W}[a]_l$ is also independent of $a$, 
and induces $r$ componentwise.
\begin{lemma}\label{lem:P}
\begin{align}
&\sigma^\Delta \Pb^{-1}s
= P^{-1} \;\; \hbox{ on }P(\coprod_{b \in J}W[b]_l),
\label{eq:sig1}\\
&\sigma^\Delta P^{-1}s
= P^{-1}Q \;\; \hbox{ on }\Pb(\coprod_{b \in J}W[b]_l),
\label{eq:sig2}
\end{align}
where $Q$ is the one for $a=-1$ case in (\ref{eq:cell})-(\ref{eq:Qdef}).
\end{lemma}
The essential task in proving Theorem \ref{th:main} is to establish
\begin{proposition}\label{pr:main}
For any $a \in J$ and $1 \le t \le 2n-2$, the action of 
the Weyl group operator $S_{i_t}$ in (\ref{eq:sww}) 
is expressed as follows:
\begin{alignat}{2}
S_{i_t} &= 
P^{-1}s_{i_t}\cdots s_{i_1}K_{j_t}
s_{i_1}\cdots s_{i_{t-1}}P & &
\hbox{if} \;\; K'_{j_t},\cdots, K'_{j_1} \hbox{ do not contain } Q,
\label{eq:main1}\\
&= \Pb^{-1}s_{i_t}\cdots s_{i_1}K_{j_t}Q
s_{i_1}\cdots s_{i_{t-1}}P & \quad
&\hbox{if} \;\; K'_{j_t} = K_{j_t}Q,\label{eq:main2}\\
&= \Pb^{-1}s_{i_t}\cdots s_{i_1}K_{j_t}
s_{i_1}\cdots s_{i_{t-1}}\Pb & 
&\hbox{ otherwise}.\label{eq:main3}
\end{alignat}
\end{proposition}
In fact, substituting (\ref{eq:main1})--(\ref{eq:main3}) 
into (\ref{eq:calTexp}) and 
reducing the result by Lemma \ref{lem:P}, $s_i^2=id$ and 
$s_{i_{2n-2}}\cdots s_{i_1} = s$, 
one gets (\ref{eq:Texp}).
Note that $a=-1$ is the unique 
case where (\ref{eq:main2}) and (\ref{eq:main3}) become irrelevant 
for all $1 \le t \le 2n-2$.
Accordingly the relation (\ref{eq:sig2}) is used only for $a = -1$ and 
the other cases are dealt with (\ref{eq:sig1}). 
A proof of Proposition \ref{pr:main} is 
included in Appendix \ref{app:b}.

\begin{remark}\label{rem:aho}
The case (\ref{eq:main2}) is relevant to $S_0$ only.
The operator $S_1$ is always expressible by (\ref{eq:main1}).
\end{remark}

\begin{example}\label{ex:t1}
For $D^{(1)}_4$, formulas (\ref{eq:main1})--(\ref{eq:main3}) read
\begin{alignat*}{2}
&a=1 & &a=2\\
&S_0 = \Pb^{-1}s_0K_{-2}QP, &
&S_1 = P^{-1}s_1K_{1}P\\
&S_2 = \Pb^{-1}s_2s_0K_{-3}s_0\Pb, &
&S_0 = \Pb^{-1}s_0s_1K_{-1}Qs_1P, \\
&S_3 = \Pb^{-1}s_3s_2s_0K_{-4}s_0s_2\Pb,&
&S_2 = \Pb^{-1}s_2s_0s_1K_{-3}s_1s_0\Pb,\\
&S_4 = \Pb^{-1}s_4s_3s_2s_0K_{4}s_0s_2s_3\Pb,&
&S_3 = \Pb^{-1}s_3s_2s_0s_1K_{-4}s_1s_0s_2\Pb,\\
&S_2 = \Pb^{-1}s_2s_4s_3s_2s_0K_{3}s_0s_2s_3s_4\Pb,&
&S_4 = \Pb^{-1}s_4s_3s_2s_0s_1K_{4}s_1s_0s_2s_3\Pb,\\
&S_0 = \Pb^{-1}s_0s_2s_4s_3s_2s_0K_{2}s_0s_2s_3s_4s_2\Pb,& \qquad
&S_2 = \Pb^{-1}s_2s_4s_3s_2s_0s_1K_{3}s_1s_0s_2s_3s_4\Pb,
\end{alignat*}
\begin{alignat*}{2}
&a=3 & &a=4\\
&S_2 = P^{-1}s_2K_{2}P, &
&S_3 = P^{-1}s_3K_{3}P\\
&S_1 = P^{-1}s_1s_2K_{1}s_2P, &
&S_2 = P^{-1}s_2s_3K_{2}s_3P, \\
&S_0 = \Pb^{-1}s_0s_1s_2K_{-1}Qs_2s_1P,&
&S_1 = P^{-1}s_1s_2s_3K_{1}s_3s_2P,\\
&S_2 = \Pb^{-1}s_2s_0s_1s_2K_{-2}s_2s_1s_0\Pb,&
&S_0 = \Pb^{-1}s_0s_1s_2s_3K_{-1}Qs_3s_2s_1P,\\
&S_3 = \Pb^{-1}s_3s_2s_0s_1s_2K_{-4}s_2s_1s_0s_2\Pb,&
&S_2 = \Pb^{-1}s_2s_0s_1s_2s_3K_{-2}s_3s_2s_1s_0\Pb,\\
&S_4 = \Pb^{-1}s_4s_3s_2s_0s_1s_2K_{4}s_2s_1s_0s_2s_3\Pb,& \qquad
&S_3 = \Pb^{-1}s_3s_2s_0s_1s_2s_3K_{-3}s_3s_2s_1s_0s_2\Pb,
\end{alignat*}
\begin{alignat*}{2}
&a=-4 & &a=-3\\
&S_4 = P^{-1}s_4K_{3}P, &
&S_3 = P^{-1}s_3K_{-4}P\\
&S_2 = P^{-1}s_2s_4K_{2}s_4P, &
&S_4 = P^{-1}s_4s_3K_{4}s_3P, \\
&S_1 = P^{-1}s_1s_2s_4K_{1}s_4s_2P,&
&S_2 = P^{-1}s_2s_4s_3K_{2}s_3s_4P,\\
&S_0 = \Pb^{-1}s_0s_1s_2s_4K_{-1}Qs_4s_2s_1P,&
&S_1 = P^{-1}s_1s_2s_4s_3K_{1}s_3s_4s_2P,\\
&S_2 = \Pb^{-1}s_2s_0s_1s_2s_4K_{-2}s_4s_2s_1s_0\Pb,&
&S_0 = \Pb^{-1}s_0s_1s_2s_4s_3K_{-1}Qs_3s_4s_2s_1P,\\
&S_4 = \Pb^{-1}s_4s_2s_0s_1s_2s_4K_{-3}s_4s_2s_1s_0s_2\Pb,& \qquad
&S_2 = \Pb^{-1}s_2s_0s_1s_2s_4s_3K_{-2}s_3s_4s_2s_1s_0\Pb,
\end{alignat*}
\begin{alignat*}{2}
&a=-2 & &a=-1\\
&S_2 = P^{-1}s_2K_{-3}P, &
&S_1 = P^{-1}s_1K_{-2}P\\
&S_3 = P^{-1}s_3s_2K_{-4}s_2P, &
&S_2 = P^{-1}s_2s_1K_{-3}s_1P, \\
&S_4 = P^{-1}s_4s_3s_2K_{4}s_2s_3P,&
&S_3 = P^{-1}s_3s_2s_1K_{-4}s_1s_2P,\\
&S_2 = P^{-1}s_2s_4s_3s_2K_{3}s_2s_3s_4P,&
&S_4 = P^{-1}s_4s_3s_2s_1K_{4}s_1s_2s_3P,\\
&S_1 = P^{-1}s_1s_2s_4s_3s_2K_{1}s_2s_3s_4s_2P,&
&S_2 = P^{-1}s_2s_4s_3s_2s_1K_{3}s_1s_2s_3s_4P,\\
&S_0 = \Pb^{-1}s_0s_1s_2s_4s_3s_2K_{-1}Qs_2s_3s_4s_2s_1P,& \qquad
&S_1 = P^{-1}s_1s_2s_4s_3s_2s_1K_{2}s_1s_2s_3s_4s_2P.
\end{alignat*}
For instance in $a=1$ case, (\ref{eq:sww}) implies that 
the first $S_0$ acts as $W[1]_l \rightarrow W[-2]_l$ whereas
the last $S_0$ does as $W[2]_l \rightarrow W[-1]_l$.
Such a distinction has not been exhibited in the left hand sides.
\end{example}

\begin{example}\label{ex:t3}
The commutative diagram corresponding to $a=3$ case 
in Example \ref{ex:t1}.
\begin{equation*}
\begin{CD}
\ldots \vert \bar{1}\bar{2}\bar{3}\bar{4}4321\vert\ldots 
@<{id}<<
\ldots \vert \bar{1}\bar{2}\bar{3}\bar{4}4321\vert\ldots
@<{P}<< W[3]_l\\
@V{K_2}VV & &  @VV{S_2}V \\
\ldots \vert \bar{1}\bar{3}\bar{2}\bar{4}4231\vert\ldots 
@<{s_2}<<
\ldots \vert \bar{1}\bar{2}\bar{3}\bar{4}4321\vert\ldots
@<{P}<< W[2]_l\\
@V{K_1}VV & &  @VV{S_1}V \\
\ldots \vert \bar{3}\bar{1}\bar{2}\bar{4}4213\vert\ldots 
@<{s_2s_1}<<
\ldots \vert \bar{1}\bar{2}\bar{3}\bar{4}4321\vert\ldots
@<{P}<< W[1]_l\\
@V{Q}VV & & & \\
\ldots \vert 3\bar{1}\bar{2}\bar{4}421{\bar 3}\vert\ldots 
& & & & @VV{S_0}V \\
@V{K_{-1}}VV & & & \\
\ldots \vert \bar{1}3\bar{2}\bar{4}42\bar{3}1\vert\ldots 
@<{s_2s_1s_0}<<
\ldots \vert 1\bar{2}\bar{3}\bar{4}432\bar{1}\vert\ldots
@<{\bar{P}}<< W[-2]_l\\
@V{K_{-2}}VV & &  @VV{S_2}V \\
\ldots \vert \bar{1}\bar{2}3\bar{4}4\bar{3}21\vert\ldots 
@<{s_2s_1s_0s_2}<<
\ldots \vert 1\bar{2}\bar{3}\bar{4}432\bar{1}\vert\ldots
@<{\bar{P}}<< W[-3]_l\\
@V{K_{-4}}VV & &  @VV{S_3}V \\
\ldots \vert \bar{1}\bar{2}\bar{4}3\bar{3}421\vert^{\small \diamondsuit}\ldots 
@<{s_2s_1s_0s_2s_3}<<
\ldots \vert 1\bar{2}\bar{3}\bar{4}432\bar{1}\vert\ldots
@<{\bar{P}}<< W[-4]_l\\
@V{K_{4}}VV & &  @VV{S_4}V \\
\ldots \vert \bar{1}\bar{2}\bar{3}\bar{4}4321\vert\ldots 
@<{s^{-1}}<<
\ldots \vert 1\bar{2}\bar{3}\bar{4}432\bar{1}\vert\ldots
@<{\bar{P}}<< W[3]_l
\end{CD}
\end{equation*}
Here, $\ldots \vert 3\bar{1}\bar{2}\bar{4}421{\bar 3}\vert\ldots$
for example indicates that in all the cells 
the letters {}from $J=\{\pm 1,\ldots, \pm 4\}$ are ordered as
\begin{equation*}
3...3-\!\!1...\!-\!\!1-\!\!2...\!-\!\!2-\!\!4...\!-\!\!4 \,4... 4 
\,2 ... 2\, 
1 ... 1 -\!\!3 ... \!-\!\!3.
\end{equation*}
In the case with (resp. without) the symbol 
{\small ${\small \diamondsuit}$}, 
$3\bar{3}$ may be written as $\bar{3}3$
(resp. $\bar{4}4$ may be written as $4\bar{4}$)
because the cells do not contain these letters simultaneously.
\end{example}

\section{Soliton}\label{sec:soliton}

Let us present the configurations that behave as solitons, namely, 
some special patterns of local states that undergo only translation 
under the time evolution ${\mathcal T}_a = T_a$.
We concentrate on the basic $D^{(1)}_n$-automaton 
but treat the general boundary condition labeled with 
an arbitrary $a \in J$.
The case $a=1$ was studied in \cite{HKT1}.
Solitons in the inhomogeneous $D^{(1)}_n$-automaton 
can then be understood along the argument in section 5 of \cite{HKOTY}.
See also section III in  \cite{HHIKTT}.
Note that the operators $P$ and $Q$ become trivial for the basic 
automaton, hence they can be dropped in Definition \ref{def:T}.

Fix $a \in J$ which specifies the boundary condition as in (\ref{eq:W}).
Let $i_1, \ldots, i_{2n-2} \in \{0,1,\ldots, n\}$ be any sequence 
such that $\varepsilon_{i_t}(r_{i_{t-1}}\cdots r_{i_{1}}(a)) > 0$ for 
all $1 \le t \le 2n-2$ and 
$r_{i_{2n-2}}\cdots r_{i_{1}}(a) = r(a)$.
($r$ is defined after (\ref{eq:sww}).)
Such a sequence exists for any $a$ and 
corresponds to a path on the crystal graph (Figure \ref{fig:B})
going {}from $a$ to $r(a)$ backward the arrows.
Set $a_t = r_{i_{t-1}}\cdots r_{i_{1}}(a)$. 
By the definition $a_{2n-2} = r(a)$ holds. 
There is no $a$ and exactly one $-a$ in the array 
$a_1, a_2, \ldots, a_{2n-3}$.
\begin{proposition}\label{pr:soliton}
Consider the following state of the basic $D^{(1)}_n$-automaton:
\begin{equation}\label{eq:pattern}
\ldots a\, a\, \underline{\overbrace{a_1\ldots a_1}^{y_1}
\overbrace{a_2\ldots a_2}^{y_2}\ldots
\overbrace{a_{2n-3}\ldots a_{2n-3}}^{y_{2n-3}}
\overbrace{a_{2n-2}\ldots a_{2n-2}}^{y_{2n-2}}}\, a\, a\,  \ldots,
\end{equation}
where $\ldots$ on the both ends stands for the array of $a$ only, and 
$y_1,\ldots, y_{2n-2}$ are non-negative integers satisfying 
\begin{equation*}
y_{2n-2}=
\begin{cases} 0 & \hbox{ if }\;  a \in \{ \pm 1, \pm n \}\;\; 
(i.e., r(a) \neq a),\\
y_p & \hbox{otherwise},
\end{cases}
\end{equation*}
where $p$ is determined by $a_p = -a$.
Under the time evolution ${\mathcal T}_a = T_a$, the underlined pattern
is shifted to the right by $y_1 + \cdots + y_{2n-2}$ lattice units.
\end{proposition}
\begin{proof}
The case $a=1$ was proved in \cite{HKT1}.
{}From Definition \ref{def:calT}, the time evolutions under 
the other boundary conditions are connected by 
a similarity transformation as  
${\mathcal T}_a = w^{-1}{\mathcal T}_1w$,  
where $w$ is an element of the affine Weyl group. 
The pattern (\ref{eq:pattern}) is obtained by 
applying $w^{-1}$ to the soliton states \cite{HKT1} for $a=1$.
\end{proof}
We call the underlined pattern a soliton with 
amplitude $y_1 + \cdots + y_{2n-2}$. 
Proposition \ref{pr:soliton} states that the 
amplitude coincides with the velocity of solitons.
Note that for $a \neq \pm 1, \pm n$, the `empty box' $a$ is 
regarded also as a part of a soliton.
To include them in the right end of solitons is just a convention.
Proposition \ref{pr:soliton} is equally valid if a soliton is 
defined, instead of (\ref{eq:pattern}), as 
$\ldots \underline{\overbrace{a...a}^{\alpha}a_1,\ldots a_{2n-3}
\overbrace{a...a}^{\beta}}\ldots$ for any $\alpha$ and $\beta$ such that 
$\alpha + \beta = y_{2n-2}$.
If there are more than one solitons, they 
move independently as described in Proposition \ref{pr:soliton}
as long as they are sufficiently separated {}from each other.
As for their scattering theory, see \cite{HKOTY}.

Let us explicitly list the possible choices of the array 
$a_1, a_2, \ldots, a_{2n-3}$ for $a \in \{\pm 1, \pm n\}$ and 
$a_1, a_2, \ldots, a_{2n-2}$ for 
$a \in \{ \pm 2, \ldots, \pm(n-1) \}$.
\begin{align*}
&-\!2,-3,\ldots -\!n\!+\!1, {n \brace -n}, 
n\!-\!1, \ldots, 3, 2
\quad \hbox{ for }\; a = \pm 1,\\
&n\!-\!1, \ldots, 2, {1 \brace -1}, -2, \ldots, 
-n\!+\!1\quad\hbox{ for }\;  a = \pm n,\\
&a\!-\!1, a\!-\!2, \ldots, 2, {1 \brace -1}, -2, \ldots, -n\!+\!1,
{n \brace -n}, n\!-\!1, \ldots, a\!+\!1, a
\;\hbox{ for }\;  2 \le a \le n\!-\!1,\\
&a\!-\!1, a\!-\!2, \ldots, -n\!+\!1, {n \brace -n},
n\!-\!1, \ldots, 2, {1 \brace -1},-2, \ldots,  
a\!+\!1, a
\;\hbox{ for }\,-\!n\!+\!1 \le a \le -2.
\end{align*}
In the first line, the 
symbol ${n \brace -n}$ means that both $T_1$ and $T_{-1}$ 
admit two arrays 
corresponding to the choices of $n$ or $-n$.
The second line is similar.
In the last two lines, each $T_a \;
(a \in \{ \pm 2, \ldots, \pm(n-1) \})$ admits 
four arrays.
\begin{example}\label{ex:soliton1}
Under the time evolution $T_2 = K_3K_4K_{-4}K_{-3}K_{-1}K_1$ 
in  the basic $D^{(1)}_4$-automaton, one has
(to save the space we write $\bar 3$ to mean $-3$ etc.) 
\begin{align*}
&\ldots 2\;2\;\underline{1\;\bar2\;\bar2\;\bar3\;4\;4\;3\;2\;2}
\;2\;2\;2\;2\;2\;2\;2\;2\;2\;2\;2\;2\ldots\\
\overset{K_1}{\longmapsto}\;\;
&\ldots 2\;2\;2\;\bar1\;\bar1\;\bar3
\;4\;4\;3\;1\;1\;1\;2\;2\;2\;2\;2\;2\;2\;2\;2\;2\;2\ldots\\
\overset{K_{-1}}{\longmapsto}\;\;
&\ldots 2\;2\;2\;2\;2\;\bar3\;4\;4\;3\;\bar2\;\bar2
\;1\;2\;2\;2\;2\;2\;2\;2\;2\;2\;2\;2\ldots\\
\overset{K_{-3}}{\longmapsto}\;\;
&\ldots 2\;2\;2\;2\;2\;2\;4\;4\;\bar2\;3\;3\;1\;\bar3\;\bar3
\;2\;2\;2\;2\;2\;2\;2\;2\;2\ldots\\
\overset{K_{-4}}{\longmapsto}\;\;
&\ldots2\;2\;2\;2\;2\;2\;4\;4\;4\;3\;3\;1\;
\bar3\;\bar3\;\bar4\;2\;2\;2\;2\;2\;2\;2\;2\ldots\\
\overset{K_{4}}{\longmapsto}\;\; 
&\ldots 2\;2\;2\;2\;2\;2\;2\;2\;2\;3\;3\;1\;
\bar3\;\bar3\;\bar2\;4\;4\;2\;2\;2\;2\;2\;2\ldots\\
\overset{K_{3}}{\longmapsto}\;\; 
&\ldots 2\;2\;2\;2\;2\;2\;2\;2\;2\;2\;2\;\underline{1\;
\bar2\;\bar2\;\bar3\;4\;4\;3\;2\;2}\;2\;2\;2\ldots
\end{align*}
A soliton with amplitude 9 (underlined) 
is shifted to the right by 9 lattice units.
\end{example}

\appendix
\section{\mathversion{bold}Crystal structure 
of $B_k$ for $D^{(1)}_n$}\label{app:a}

Here we quote the $D^{(1)}_n$ crystal structure of $B_k$
\cite{KKM} in a form adapted to section \ref{sec:5}.
In this paper we write the Kashiwara operators 
${\tilde e}_i$ and ${\tilde f}_i$ ($0 \le i \le n$)
as $e_i$ and $f_i$, respectively.
Moreover we employ the convention 
$e^q_i = f^{-q}_i$ for $q < 0$.
Let $\vec x = (x_1,\ldots, x_n,x_{-n},\ldots, x_{-1})$ be an
element of the $D^{(1)}_n$-crystal as described in Table \ref{tab:B}.
For $q \in \Z$, we set $e_i^q(\vec x) = \vec x'$. We assume that $\vec x' \neq 0$ and let 
$\Delta x_j = x'_j - x_j$ denote the change of the coordinates ($j \in J$).
For each $0 \le i \le n$, all the $\Delta x_j$'s are zero except the 
following:
\begin{alignat}{2}
&i=0& \;
&\begin{cases}
\Delta x_{-1} = -\Delta x_{2} = 
(x_{2}-x_{-2})_+ - (x_{2}-x_{-2}-q)_+,&\\
\Delta x_1 = -\Delta x_{-2} = 
(x_{-2}-x_{2})_+ - (x_{-2}-x_{2}+q)_+,&
\end{cases}\\
&1 \le i \le n\!-\!1& 
&\begin{cases}
\Delta x_{i} = -\Delta x_{i+1} = 
(x_{i+1}-x_{-i-1})_+ - (x_{i+1}-x_{-i-1}-q)_+,&\\
\Delta x_{-i} = -\Delta x_{-i-1} = 
(x_{-i-1}-x_{i+1})_+ - (x_{-i-1}-x_{i+1}+q)_+,&
\end{cases}\label{eq:henka}\\
&i=n& \quad 
&\begin{cases}
\Delta x_{n-1} = -\Delta x_{-n} = 
x_{-n} - (x_{-n}-x_{n}-q)_+,&\\
\Delta x_{-n+1} = -\Delta x_{n} = 
x_{n} - (x_{n}-x_{-n}+q)_+.&
\end{cases}
\end{alignat}
Here we have used the symbol $(z)_+ = \max(z,0)$.

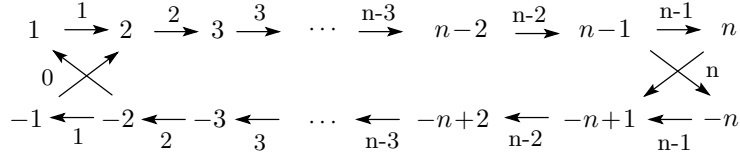
\begin{figure}[h]
\unitlength 0.1in
\begin{picture}( 39.1100,  7.2000)( 10.6000,-15.100)
\put(18.4000,-9.2000){\makebox(0,0){1}}%
\put(40.7200,-9.2000){\makebox(0,0){$n\!-\!2$}}%
\put(54.7600,-9.2000){\makebox(0,0){$n$}}%
\put(48.2800,-9.2000){\makebox(0,0){$n\!-\!1$}}%
\put(28.0000,-9.2000){\makebox(0,0){3}}%
\put(23.2000,-9.2000){\makebox(0,0){2}}%
\put(47.9600,-13.9000){\makebox(0,0){$-n\!+\!1$}}%
\put(54.3200,-13.9000){\makebox(0,0){$-n$}}%
\put(40.2800,-13.9000){\makebox(0,0){$-n\!+\!2$}}%
\put(27.5600,-13.9000){\makebox(0,0){$-3$}}%
\put(22.7600,-13.9000){\makebox(0,0){$-2$}}%
\put(17.9600,-13.9000){\makebox(0,0){$-1$}}%
%
\special{pn 8}%
\special{pa 2000 920}%
\special{pa 2220 920}%
\special{fp}%
\special{sh 1}%
\special{pa 2220 920}%
\special{pa 2154 900}%
\special{pa 2168 920}%
\special{pa 2154 940}%
\special{pa 2220 920}%
\special{fp}%
\put(20.9000,-8.2000){\makebox(0,0){{\small 1}}}%
%
\special{pn 8}%
\special{pa 2470 930}%
\special{pa 2690 930}%
\special{fp}%
\special{sh 1}%
\special{pa 2690 930}%
\special{pa 2624 910}%
\special{pa 2638 930}%
\special{pa 2624 950}%
\special{pa 2690 930}%
\special{fp}%
%
\special{pn 8}%
\special{pa 2900 930}%
\special{pa 3120 930}%
\special{fp}%
\special{sh 1}%
\special{pa 3120 930}%
\special{pa 3054 910}%
\special{pa 3068 930}%
\special{pa 3054 950}%
\special{pa 3120 930}%
\special{fp}%
%
\special{pn 8}%
\special{pa 3540 930}%
\special{pa 3760 930}%
\special{fp}%
\special{sh 1}%
\special{pa 3760 930}%
\special{pa 3694 910}%
\special{pa 3708 930}%
\special{pa 3694 950}%
\special{pa 3760 930}%
\special{fp}%
%
\special{pn 8}%
\special{pa 5100 920}%
\special{pa 5320 920}%
\special{fp}%
\special{sh 1}%
\special{pa 5320 920}%
\special{pa 5254 900}%
\special{pa 5268 920}%
\special{pa 5254 940}%
\special{pa 5320 920}%
\special{fp}%
%
\special{pn 8}%
\special{pa 4360 940}%
\special{pa 4580 940}%
\special{fp}%
\special{sh 1}%
\special{pa 4580 940}%
\special{pa 4514 920}%
\special{pa 4528 940}%
\special{pa 4514 960}%
\special{pa 4580 940}%
\special{fp}%
%
\special{pn 8}%
\special{pa 2150 1380}%
\special{pa 1930 1380}%
\special{fp}%
\special{sh 1}%
\special{pa 1930 1380}%
\special{pa 1998 1400}%
\special{pa 1984 1380}%
\special{pa 1998 1360}%
\special{pa 1930 1380}%
\special{fp}%
%
\special{pn 8}%
\special{pa 4530 1380}%
\special{pa 4310 1380}%
\special{fp}%
\special{sh 1}%
\special{pa 4310 1380}%
\special{pa 4378 1400}%
\special{pa 4364 1380}%
\special{pa 4378 1360}%
\special{pa 4310 1380}%
\special{fp}%
\put(51.9000,-15.1000){\makebox(0,0){{\small n-1}}}%
%
\special{pn 8}%
\special{pa 5280 1390}%
\special{pa 5060 1390}%
\special{fp}%
\special{sh 1}%
\special{pa 5060 1390}%
\special{pa 5128 1410}%
\special{pa 5114 1390}%
\special{pa 5128 1370}%
\special{pa 5060 1390}%
\special{fp}%
%
\special{pn 8}%
\special{pa 2630 1390}%
\special{pa 2410 1390}%
\special{fp}%
\special{sh 1}%
\special{pa 2410 1390}%
\special{pa 2478 1410}%
\special{pa 2464 1390}%
\special{pa 2478 1370}%
\special{pa 2410 1390}%
\special{fp}%
%
\special{pn 8}%
\special{pa 3120 1390}%
\special{pa 2900 1390}%
\special{fp}%
\special{sh 1}%
\special{pa 2900 1390}%
\special{pa 2968 1410}%
\special{pa 2954 1390}%
\special{pa 2968 1370}%
\special{pa 2900 1390}%
\special{fp}%
%
\special{pn 8}%
\special{pa 3750 1390}%
\special{pa 3530 1390}%
\special{fp}%
\special{sh 1}%
\special{pa 3530 1390}%
\special{pa 3598 1410}%
\special{pa 3584 1390}%
\special{pa 3598 1370}%
\special{pa 3530 1390}%
\special{fp}%
\put(44.1000,-14.9000){\makebox(0,0){{\small n-2}}}%
\put(53.9000,-11.4000){\makebox(0,0){{\small n}}}%
\put(36.6000,-15.0000){\makebox(0,0){{\small n-3}}}%
\put(20.7000,-14.8000){\makebox(0,0){{\small 1}}}%
\put(25.3000,-15.0000){\makebox(0,0){{\small 2}}}%
\put(25.7000,-8.4000){\makebox(0,0){{\small 2}}}%
\put(30.2000,-15.1000){\makebox(0,0){{\small 3}}}%
\put(33.600,-14.100){\makebox(0,0){{\small $\cdots$}}}%
\put(30.2000,-8.3000){\makebox(0,0){{\small 3}}}%
\put(33.600,-9.2000){\makebox(0,0){{\small $\cdots$}}}%
\put(36.4000,-8.3000){\makebox(0,0){{\small n-3}}}%
\put(44.3000,-8.4000){\makebox(0,0){{\small n-2}}}%
\put(52.0000,-8.2000){\makebox(0,0){{\small n-1}}}%
\put(19.1000,-11.7000){\makebox(0,0){{\small 0}}}%
%
\special{pn 8}%
\special{pa 5066 1030}%
\special{pa 5372 1260}%
\special{fp}%
\special{sh 1}%
\special{pa 5372 1260}%
\special{pa 5330 1204}%
\special{pa 5328 1228}%
\special{pa 5306 1236}%
\special{pa 5372 1260}%
\special{fp}%
%
\special{pn 8}%
\special{pa 5342 1038}%
\special{pa 5036 1268}%
\special{fp}%
\special{sh 1}%
\special{pa 5036 1268}%
\special{pa 5102 1244}%
\special{pa 5080 1236}%
\special{pa 5078 1212}%
\special{pa 5036 1268}%
\special{fp}%
%
\special{pn 8}%
\special{pa 1970 1270}%
\special{pa 2276 1040}%
\special{fp}%
\special{sh 1}%
\special{pa 2276 1040}%
\special{pa 2212 1064}%
\special{pa 2234 1072}%
\special{pa 2236 1096}%
\special{pa 2276 1040}%
\special{fp}%
%
\special{pn 8}%
\special{pa 2248 1262}%
\special{pa 1942 1032}%
\special{fp}%
\special{sh 1}%
\special{pa 1942 1032}%
\special{pa 1982 1088}%
\special{pa 1984 1064}%
\special{pa 2006 1056}%
\special{pa 1942 1032}%
\special{fp}%
\end{picture}%
\unitlength 0.35mm
\caption{Crystal graph of $B_1$}
\label{fig:B}
\end{figure}

If one associates a node to each element $\vec x \in B_k$ and 
a figure $\vec x \overset{i}{\leftarrow} \vec x'$
to each relation $e_i(\vec x) = \vec x'$, the result is 
an oriented graph called a crystal graph, 
whose arrows carry the colors $0,1,\ldots, n$.
For $k=1$, the crystal graph of $B_1$ looks as
Figure \ref{fig:B}.
\vspace{0.1cm}

\vspace{0.1cm}
Here the elements of $B_1$ are denoted by the letter in $J$.
(See the second paragraph of section \ref{sec:3}.)

The functions $\varphi_i, \varepsilon_i: B_k \rightarrow \Z_{\ge 0}$ are 
given by 
\begin{align}
&\varphi_0(\vec x) = x_{-1}+(x_{-2}-x_2)_+,\quad 
\varepsilon_0(\vec x) = x_{1}+(x_{2}-x_{-2})_+,\\
&\varphi_i(\vec x) = x_{i}+(x_{-i-1}-x_{i+1})_+,\quad 
\varepsilon_i(\vec x) = x_{-i}+(x_{i+1}-x_{-i-1})_+\;\; 1 \le i \le n-1,\\
&\varphi_n(\vec x) = x_{n-1}+x_{n},\quad 
\varepsilon_n(\vec x) = x_{-n+1}+x_{-n}.
\end{align}
Note that $(x_j-x_{-j})_+ = x_j$ for $j=\pm n$ 
due to the constraint $x_nx_{-n}=0$.
For $p \in B_1$, one has the simple description 
\begin{equation}\label{eq:epdef}
\varepsilon_i(p) = \begin{cases}
1 & \hbox{ if } p ' \overset{i}{\rightarrow} p \hbox{ for some }
p' \in B_1,\\
0 & \hbox{ otherwise}.
\end{cases}
\end{equation}

The Weyl group operator is defined by $S_i = e_i^{\varepsilon_i-\varphi_i}$.
Whenever $S_i$ acts on the states $\xi \in W[a]_l$, 
it acts as $S_i = e_i^\infty$ since  $\varphi_i(\xi) = 0$ holds 
due to the boundary condition.
In such a situation, $\xi'= S_i(\xi)$ can be computed 
recursively as
\begin{align}
&\xi = (\ldots \vec \xi_g, \vec \xi_{g+1}, \ldots ),\quad 
\xi' = (\ldots \vec \xi'_g, \vec \xi'_{g+1}, \ldots ),\label{eq:rule1}\\
&\vec \xi'_g = e_i^{(\varepsilon_i(\vec \xi_g)-m_g)_+}(\vec \xi_g),\quad 
m_{g+1} = (m_g - \varepsilon_i(\vec \xi_g))_+ + \varphi_i(\vec \xi_g),
\label{eq:rule2}\\
&\vec \xi'_g = e_i^{\varepsilon_i(\vec \xi_g)}(\vec \xi_g),\quad 
m_g = 0 \;\;\hbox{ for } g \ll -1.\label{eq:rule3}
\end{align}
The $m_{g+1} \in \Z_{\ge 0}$ determined as above is equal to
$\varphi_i((\ldots, \vec \xi_{g-1}, \vec \xi_{g}))$. 

\section{\mathversion{bold}Proof of 
Proposition \ref{pr:main}}\label{app:b}

Here we exhibit the dependence of the 
operators $K_j$ and $Q$ on the boundary condition $a \in J$ 
as $K^{(a)}_j$ and $Q^{(a)}$.
By the definition it is easy to see
\begin{lemma}\label{lem:reduce}
For any $0 \le i \le n$ and $a, j \in J$, the following is valid:
\begin{align}
&s_iK^{(a)}_j= K^{(r_i(a))}_{r_i(j)}s_i,\label{eq:ks}\\
&s_iQ^{(a)} = Q^{(r_i(a))}s_i.\label{eq:qs}
\end{align}
\end{lemma}
We denote by $S_i\vert_{W[a]_l}$ the 
restriction of $S_i$ on $W[a]_l$.
\begin{lemma}\label{lem:sk}
\begin{alignat}{2}
&S_0\vert_{W[1]_l} = \Pb^{-1} s_0 K^{(1)}_{-2}Q^{(1)}P,\quad &
&S_0\vert_{W[2]_l} = \Pb^{-1} s_0 K^{(2)}_{-1}\Pb,\label{eq:s0}\\
&S_1\vert_{W[2]_l} = P^{-1} s_1 K^{(2)}_{1}P, &
&S_1\vert_{W[-1]_l} = P^{-1} s_1 K^{(-1)}_{-2}P,\label{eq:s1}\\
&S_i\vert_{W[c]_l} = P^{-1} s_i K^{(c)}_{r_i(c)}P
= \Pb^{-1} s_i K^{(c)}_{r_i(c)}\Pb && 
\qquad (i \neq 0,1),\label{eq:si}
\end{alignat}
where in (\ref{eq:si}), 
$c$ is any element of $J = B_1$ satisfying $\varepsilon_i(c) > 0$.
(For any $i \neq 0,1$, there are two such $c$, 
which are contained in the 
crystal graph as $r_i(c) \overset{i}{\longrightarrow} c$.)
\end{lemma}
\begin{proof}
Let us show 
$S_i\vert_{W[i+1]_l} = P^{-1}s_iK^{(i+1)}_iP$ for $2 \le i \le n-1$, 
which is a part of the assertion (\ref{eq:si}). 
The other cases can be verified similarly.
Obviously, the map 
$s_iK^{(i+1)}_i: \widetilde{W}[i+1]_l \rightarrow \widetilde{W}[i]_l$ 
only changes the letters 
$\pm i, \pm(i+1)$ in each cell.
So setting $\vec \xi_g = (x_1,\ldots, x_{-1}) \in B_{l_g}$, 
we only consider their array within a cell:
\begin{equation*}
\overbrace{-i,\ldots, -i}^{x_{-i}}
\overbrace{-i\!-\!1,\ldots, -i\!-\!1}^{x_{-i-1}}
\overbrace{i\!+\!1,\ldots, i\!+\!1}^{x_{i+1}}
\overbrace{i,\ldots, i}^{x_{i}},
\end{equation*}
dropping the other letters.
When $K^{(i+1)}_i$ is applied on this array according to 
the vertex diagram rule (\ref{eq:LL}), it is transformed as 

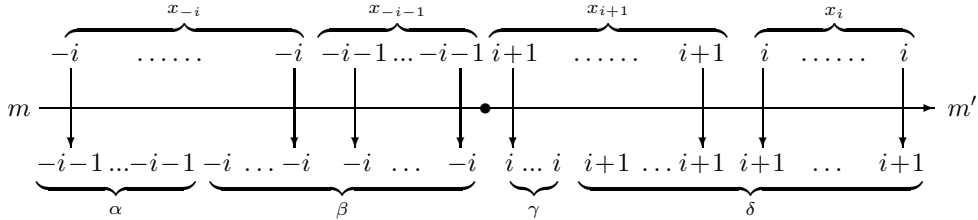
\begin{picture}(360,100)(5,-32)
\put(10,20){\vector(1,0){340}}
\put(20,46){$\overbrace{\qquad\qquad\qquad\qquad\quad}^{x_{-i}}$}
\put(22,35){\vector(0,-1){30}}\put(14,38){$-i\quad \quad\ldots \ldots$}
\put(9,-4){$-i\!-\!1\,...$}
\put(9,-7){$\underbrace{\qquad\qquad\qquad}_{\alpha}$}
\put(116,46){$\overbrace{\qquad\qquad\qquad}^{x_{-i-1}}$}
\put(44,-4){$-i\!-\!1$}
\put(72,-4){$-i\; \ldots$}
\put(75,-7){$\underbrace{\qquad\qquad\qquad\qquad\qquad}_{\beta}$}
\put(107,35){\vector(0,-1){30}}\put(99,38){$-i$}\put(102,-4){$-i$}
\put(130,35){\vector(0,-1){30}}\put(117,38){$-i\!-\!1\, ...$}
\put(125,-4){$-i\;\; \ldots$}
\put(170,35){\vector(0,-1){30}}\put(154,38){$-i\!-\!1$}\put(165,-4){$-i$}
\put(181,46){$\overbrace{\qquad\qquad\qquad\qquad\quad}^{x_{i+1}}$}
\put(190,35){\vector(0,-1){30}}\put(182,38){$i\!+\!1\quad\, \ldots\ldots $}
\put(187,-4){$i \; ...$}
\put(189,-7){$\underbrace{\quad}_{\gamma}$}
\put(205,-4){$i$}\put(217,-4){$i\!+\!1\; \ldots$}
\put(215,-7){$\underbrace{\qquad\qquad\qquad\qquad\qquad\qquad\quad}_{\delta}$}%
\put(262,35){\vector(0,-1){30}}
\put(253,38){$i\!+\!1$}
\put(253,-4){$i\!+\!1$}
\put(282,46){$\overbrace{\qquad\qquad\qquad}^{x_{i}}$}
\put(285,35){\vector(0,-1){30}}\put(284,38){$i\quad \ldots\ldots$}
\put(276,-4){$i\!+\!1\;\;\; \ldots$}
\put(338,35){\vector(0,-1){30}}\put(337,38){$i$}\put(329,-4){$i\!+\!1$}
\put(-2,17){$m$}
\put(355,17){$m'$}
\put(177.1,17.5){$\bullet$}
\end{picture}

\noindent
where we have assumed that the number 
attached to the horizontal line as in (\ref{eq:LL}) is $m$ 
for the cell in question.
Here $\alpha, \beta, \gamma, \delta$ and $m'$ are given by 
\begin{align}
&\alpha = \min(x_{-i},m), \quad
\beta = x_{-i-1} + (x_{-i}-m)_+,\\
&\gamma = \min(x_{i+1},{\tilde m}),\quad 
\delta = x_i + (x_{i+1}-{\tilde m})_+,\\
&m' = x_i + ({\tilde m}-x_{i+1})_+,\quad 
{\tilde m} = x_{-i-1} + (m-x_{-i})_+,
\end{align}
where ${\tilde m}$ is the number attached to the horizontal line at 
the position marked with the symbol $\bullet$.
Thus under $P^{-1}s_iK^{(i+1)}_iP$,  
the element $\vec \xi_g$ is transformed into 
$\vec \xi'_g = (x'_1,\ldots, x'_{-1}) \in B_{l_g}$ with 
non-trivial changes given by  
$x'_{-i} = \alpha$, $x'_{-i-1} = \beta$, 
$x'_{i+1} = \gamma$ and $x'_i = \delta$.
We are to show that this coincides with the rule 
(\ref{eq:rule1})--(\ref{eq:rule3}) for $S_i$.
In view of (\ref{eq:henka}), it is equivalent to checking 
\begin{align*}
&\delta - x_i = x_{i+1}-\gamma = (x_{i+1}-x_{-i-1})_+ - 
(x_{i+1}-x_{-i-1} - q)_+,\\
&\alpha - x_{-i} = x_{-i-1}-\beta = 
(x_{-i-1}-x_{i+1})_+ - 
(x_{-i-1}-x_{i+1} + q)_+,\\
&m' = (m-x_{-i}-(x_{i+1}-x_{-i-1})_+)_+ + 
x_i + (x_{-i-1}-x_{i+1})_+,
\end{align*}
where $q = (x_{-i}+(x_{i+1}-x_{-i-1})_+-m)_+$.
These identities can be proved straightforwardly.
\end{proof}
\begin{example}\label{ex:sk2}
For $n=4$, the relations (\ref{eq:si}) read
\begin{align}
&S_2\vert_{W[3]_l} = P^{-1}s_2K^{(3)}_2P
= \Pb^{-1}s_2K^{(3)}_2\Pb,\\
&S_2\vert_{W[-2]_l} = P^{-1}s_2K^{(-2)}_{-3}P
= \Pb^{-1}s_2K^{(-2)}_{-3}\Pb,\\
&S_3\vert_{W[4]_l} = P^{-1}s_3K^{(4)}_{3}P
= \Pb^{-1}s_3K^{(4)}_{3}\Pb,\\
&S_3\vert_{W[-3]_l} = P^{-1}s_3K^{(-3)}_{-4}P
= \Pb^{-1}s_3K^{(-3)}_{-4}\Pb,\\
&S_4\vert_{W[-4]_l} = P^{-1}s_4K^{(-4)}_{3}P
= \Pb^{-1}s_4K^{(-4)}_{3}\Pb,\\
&S_4\vert_{W[-3]_l} = P^{-1}s_4K^{(-3)}_{4}P
= \Pb^{-1}s_4K^{(-3)}_{4}\Pb.
\end{align}
\end{example}

{\em Proof of Proposition \ref{pr:main}}. 
All the relations (\ref{eq:main1})--(\ref{eq:main3}) are 
reduced to Lemma \ref{lem:sk} by means of Lemma \ref{lem:reduce}.
In fact consider the case $i_t \neq 0,1$.
Then {}from Remark \ref{rem:aho}, the identities to be shown 
are either (\ref{eq:main1}) or (\ref{eq:main3}).
With the aid of Lemma \ref{lem:reduce}, they are reduced to 
\begin{equation}\label{eq:spkp}
S_{i_t}\vert_{W[r_{i_{t-1}}\cdots r_{i_1}(a)]_l} = 
P^{-1}s_{i_t}
K^{(r_{i_{t-1}}\cdots r_{i_1}(a))}_{r_{i_{t-1}}\cdots r_{i_1}(j_t)}P
\;\hbox{ or }\;
\Pb^{-1}s_{i_t}
K^{(r_{i_{t-1}}\cdots r_{i_1}(a))}_{r_{i_{t-1}}\cdots r_{i_1}(j_t)}\Pb,
\end{equation}
where in the left hand side, we have 
exhibited the fact (\ref{eq:sww}) explicitly.
Substitute (\ref{eq:ij}) into this and set 
$r_{i_{t-1}}\cdots r_{i_1}(a)=c$ and $i_t=i$.
The result coincides with (\ref{eq:si}).
Moreover, the property (\ref{eq:iprop}) tells that 
$\varepsilon_i(c)>0$. 
Thus (\ref{eq:spkp}) is reduced to Lemma \ref{lem:sk}.
The remaining case $i_t = 0,1$ can be shown similarly by means of
(\ref{eq:s0}) and (\ref{eq:s1}).
This completes a proof of Proposition \ref{pr:main}.
\hfill $\square$

\begin{remark}\label{rem:KK}
By substituting (\ref{eq:si}) into 
$S_{n-1}S_n = S_nS_{n-1}$ and applying Lemma \ref{lem:reduce}, one can show
$K^{(a)}_nK^{(a)}_{-n} = K^{(a)}_{-n}K^{(a)}_{n}$ for $a \neq \pm n$.
Similarly $S_0S_1 = S_1S_0$ leads to 
$Q^{(1)}K^{(a)}_{-1}Q^{(a)}K^{(a)}_1 = 
K^{(a)}_1Q^{(a)}K^{(a)}_{-1}Q^{(1)}$ for $a \neq \pm 1$.
\end{remark}

Finally we include a lemma concerning 
the operator $K_{-a}$ appearing in the 
$A^{(2)}_{2n-3}$-automaton and the 
vertex diagram (\ref{eq:batt2}).
\begin{lemma}\label{lem:k-a}
If $x_n = x_{-n} = 0$ for all the local states 
$\vec x = (x_1,\ldots, x_{-1})$ in the $D^{(1)}_n$-automaton,
the composition $K_nK_{-n}$ appearing in $T_a \, (a \neq \pm n)$ 
is equal to $K_{-a}$ defined around (\ref{eq:batt2}).
\end{lemma}
\begin{proof}
{}From the remark made after Example \ref{ex:Td4},
the $K_nK_{-n}$ always acts on the cells of the form
$\ldots \overbrace{a...a}^\alpha
\overbrace{-a...-\!a}^\beta \ldots$, where 
the $\ldots$ parts on the both ends do not contain 
the letters $\pm a, \pm n$ and are kept unchanged.
Therefore by dropping them, the action of $K_nK_{-n}$ on a cell 
can be computed with the following diagram:

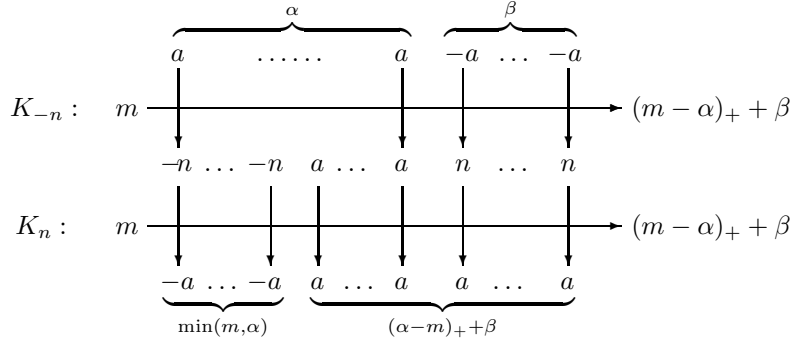
\begin{picture}(360,138)(-55,-73)
\put(10,20){\vector(1,0){180}}
\put(20,46){$\overbrace{\qquad\qquad\qquad\qquad\quad}^{\alpha}$}
\put(22,35){\vector(0,-1){30}}\put(22,-10){\vector(0,-1){30}}
\put(19,38){$a\;\;\quad \quad\ldots \ldots$}
\put(15,-4){$-\!n \; \ldots$}\put(15,-49){$-a \; \ldots$}
\put(123,46){$\overbrace{\quad\qquad\qquad}^{\beta}$}
\put(48,-4){$-n$}\put(48,-49){$-a$}
\put(57,-10){\vector(0,-1){30}}
\put(72,-4){$a\; \ldots$}\put(72,-49){$a \; \;\ldots$}
\put(75,-10){\vector(0,-1){30}}
\put(107,35){\vector(0,-1){30}}\put(107,-10){\vector(0,-1){30}}
\put(104,38){$a$}\put(104,-4){$a$}\put(104,-49){$a$}
\put(130,35){\vector(0,-1){30}}\put(130,-10){\vector(0,-1){30}}
\put(123,38){$-a\;\; \ldots $}
\put(127,-4){$n\;\;\; \ldots$}\put(127,-49){$a \;\;\, \ldots$}
\put(170,35){\vector(0,-1){30}}\put(170,-10){\vector(0,-1){30}}
\put(162,38){$-a$}\put(167,-4){$n$}\put(167,-49){$a$}
\put(16,-51){$\underbrace{\qquad\qquad\;\;}_{\min(m,\alpha)}$}
\put(72,-51){$\underbrace{
\qquad\qquad\qquad\qquad\quad\quad}_{(\alpha-m)_+ + \beta}$}
\multiput(0,0)(0,-45){2}
{\put(10,20){\vector(1,0){180}}
\put(-2,17){$m$}\put(194,17)
{$(m-\alpha)_+ + \beta$}}
\put(-42,17){$K_{-n}:$}\put(-40,-28){$K_{n}:$}
\end{picture}

\noindent
One finds that the letters $\pm n$ are indeed confined within the 
intermediate state between $K_n$ and $K_{-n}$.
Moreover, the composition $K_nK_{-n}$ induces the same 
transformation as $K_{-a}$ defined by (\ref{eq:batt2}).
\end{proof}

\vspace{0.3cm}\noindent
{\bf Acknowledgements} \hspace{0.1cm}
A.K. is partially supported by Grand-in-Aid for Scientific 
Research JSPS No.15540363.

\end{document}